\documentclass[proceedings]{JHEP37}
\usepackage{eurosym}
\usepackage{amssymb}
\usepackage{amsfonts}
\usepackage{amsmath,epsfig}
\usepackage{graphicx}

\newbox\mybox

\newcommand\fverb{\setbox\mybox=\hbox\bgroup\verb}
\newcommand\fverbdo{\egroup\medskip\noindent\fbox{\unhbox\mybox}\ }
\newcommand\fverbit{\egroup\item[\fbox{\unhbox\mybox}]}
\conference{Degenerate multi-solitons in the sine-Gordon equation}
\abstract{We construct various types of degenerate multi-soliton and multi-breather solutions for the sine-Gordon equation based on B\"{a}cklund transformations,
Darboux-Crum transformations and Hirota's direct method. We compare the different solution procedures and study the properties of the solutions. Many of them exhibit a compound like behaviour 
on a small timescale, but their individual one-soliton constituents separate for large time. Exceptions are degenerate cnoidal kink solutions that we construct via inverse scattering
from shifted Lam\'{e} potentials. These type of solutions have constant speed and do not display any time-delay. We analyse the asymptotic behaviour of the solutions and 
compute explicit analytic expressions for time-dependent displacements between the individual one-soliton constituents for any number of degeneracies. When expressed in terms of the soliton 
speed and spectral parameter the expression found is of the same generic form as the one formerly found for the Korteweg de-Vries equation.}

\title{Degenerate multi-solitons in the sine-Gordon equation}
\author{Julia Cen$^\bullet$, Francisco Correa$^\circ$ and Andreas Fring$%
^\bullet$ \\
$\bullet$ Department of Mathematics, City, University of London,\\
$\,\,$ Northampton Square, London EC1V 0HB, UK \\
$\circ$ Instituto de Ciencias F{\'{\i}}sicas y Matem{\'{a}}ticas,
Universidad Austral de Chile, \\
$\,\,$ Casilla 567, Valdivia, Chile\\
E-mail: julia.cen.1@city.ac.uk, francisco.correa@uach.cl, a.fring@city.ac.uk}

\input{tcilatex}
\begin{document}

\section{Introduction}

Previously \cite{CenFring,CorreaFring,cen2016time} we constructed compound
soliton solutions to the Korteweg de-Vries (KdV) equation that were composed
of a fixed number of one-soliton solutions. These solutions have the
interesting property that in a large regime at a small timescale the
one-solitons constituents travel simultaneously at the same speed and with
the same amplitude. Due to this property the collection of them could be
regarded as an almost stable compound. In this regime the solutions behave
similar to the famous tidal bore phenomenon which consists of multiple wave
amplitudes of heights up to several meters traveling jointly upstream a
river covering distances of up to several hundred kilometers, see e.g. \cite%
{chanson2012tidal}. However, the solutions we obtained display different
behaviour at different timescales. At very large time the individual
one-soliton constituents separate from each other with a time-dependent
displacement, which could be computed exactly in closed analytical form for
any number of one-solitons contained in the solution. For the nonlinear Schr%
\"{o}dinger equation degenerate solutions were recently studied in \cite%
{li2017degenerate}.

Technically these solutions were obtained by regularizing well-known
multi-soliton solutions when considering complex $\mathcal{PT}$-symmetric
solutions with real energies. This complexification was necessary as the
standard limit of one velocity coinciding with another leads to cusp type
solutions tending to infinity. The procedure was made conceptually possible
as the value of the overall energy was shown to remain real maintaining the
same value.

Here we argue that for the sine-Gordon equation such type of compound
multi-kink solutions also exist and the limit leading to degeneracy can be
carried out even for the real valued solutions. We study here the
sine-Gordon equation in light-cone variables%
\begin{equation}
\phi _{xt}=\sin \phi ,  \label{SG}
\end{equation}%
which we denote by $x$, $t$ and are related the original time and coordinate 
$T$, $X$, respectively, as $x=(X+T)/2~$\ and $t=(X-T)/2$. Non-degenerate
solutions to the sine-Gordon equations (\ref{SG}) are known for a long time 
\cite{HirotaSG}. A convenient compact expression may be found, when
transformed to light-cone variables, for instance in \cite{caudrey1973}, 
\begin{equation}
\phi _{\alpha _{1}\alpha _{2}\ldots \alpha _{n}}(x,t)=\arccos \left\{ 1-2 
\left[ \ln \left( \det M\right) \right] _{xt}\right\}  \label{oldsol}
\end{equation}%
with 
\begin{equation}
M_{ij}=\frac{(\alpha _{i}\alpha _{j})^{1/2}}{\alpha _{i}+\alpha _{j}}\left[
e^{\eta _{i}}+(-1)^{i+j}e^{-\eta _{j}}\right] ,~~~~~~~~\eta _{i}=x\alpha
_{i}+t/\alpha _{i}+\mu _{i},~~
\end{equation}%
and constants $\alpha _{i}$, $\mu _{i}$, $\alpha _{i}\neq \alpha _{j}$ for $%
i,j=1,\ldots ,n$. The solution (\ref{oldsol}) demonstrates the problem that
arises when one tries to obtain degenerate solutions from the previous
expressions by naively setting the parameters $\alpha _{i}=\alpha _{j}$, as
then the determinant of $M$ evidently vanishes for $n>3$ or just a constant
for $n=2$. Different versions, all plagued by a similar problem, may be
found in \cite{HirotaSG,nimmo1984,barashenkov1987}. We abbreviate solutions
with a $m$-fold degeneracy in $\alpha $ as $\phi _{m\alpha \alpha
_{m+1}\ldots \alpha _{n}}(x,t)$, i.e. this denotes an $n$-soliton solution
with $\alpha _{1}=$ $\alpha _{2}=\ldots =$ $\alpha _{m}=\alpha $.

As is well known \cite{rubinstein,jackiw,fring1994vertex}, the scattering of
one-solitons within standard multi-soliton solutions leads to displacements
in space or depending on the type of interaction delays or advances in time,
which can be used to compute quantum mechanical scattering matrices in
semi-classical approximations \cite{wignereisen}. In \cite{cen2016time} we
observed for the degenerate KdV-solutions that these displacements become
time-dependent. Here we find a similar asymptotic behaviour for the majority
of our solutions and notice that the expressions for the delays take on a
universal model independent form when expressed in terms of the soliton
speed $v$ and the constant spectral parameter $\alpha $. For a $N$-soliton
or $N$-kink solution $\phi _{Na}~$with $N$ parameterized as $N=2n+1-\kappa $
we find the universal formula for the time-dependent displacements between
the different one-soliton contributions 
\begin{equation}
\Delta _{n,\ell ,\kappa }(t):=\frac{1}{\alpha }\ln \left[ \frac{(n-\ell )!}{%
(n+\ell -\kappa )!}\left( 4\alpha \left\vert tv\right\vert \right) ^{2\ell
-\kappa }\right] \text{\thinspace \thinspace }~n=1,\ldots ;~\ell =\kappa
,\ldots ,n;~\kappa =0,1,  \label{uni}
\end{equation}%
with speeds $v=\alpha ^{2}$ and $v=-1/\alpha ^{2}$ in the KdV-solutions and
sine--Gordon kink solutions, respectively. We also construct degenerate
solutions based on cnoidal kink solutions, which have constant speed and do
not display displacements at any time.

Our manuscript is organized as follows: In sections 2, 3 and 4 we construct
degenerate multi-soliton solutions based on B\"{a}cklund transformations,
Darboux-Crum transformations and Hirota's direct method, respectively. In
section 5 we analyze the asymptotic behaviour of the solutions and compute
displacements between the different one-soliton solutions deriving our
universal formula (\ref{uni}).

\section{Degenerate multi-solitons from B\"{a}cklund transformations}

It is well-known that in general the standard superposition principle for
linear wave differential equations is replaced by model dependent functional
relations involving more than two solutions for nonlinear wave equations. In
the case of the sine-Gordon equation this equation arise from the
combination of four sets of \emph{B\"{a}cklund transformations} where each
of them relate two different solutions of (\ref{SG}), say $\phi $ and $\phi
^{\prime }$, as%
\begin{equation}
\frac{\phi _{x}+\phi _{x}^{\prime }}{2}=\frac{1}{\kappa }\sin \left[ \frac{%
\phi -\phi ^{\prime }}{2}\right] ,\qquad \text{and\qquad }\frac{\phi
_{t}-\phi _{t}^{\prime }}{2}=\kappa \sin \left[ \frac{\phi +\phi ^{\prime }}{%
2}\right] .  \label{BT}
\end{equation}%
Evidently, the two solutions in (\ref{BT}) can not be arbitrary and need to
be selected such that $\kappa $ is a constant. Taking now four solutions $%
\phi $, $\phi ^{\prime }$, $\phi ^{\prime \prime }$, $\phi ^{\prime \prime
\prime }$ and relate them pairwise as indicated in the Lamb-Bianchi diagram 
\cite{Lamb,Bianchi} in figure 1 one obtains four sets of B\"{a}cklund
transformations.

\begin{center}
\FIGURE{\epsfig{file=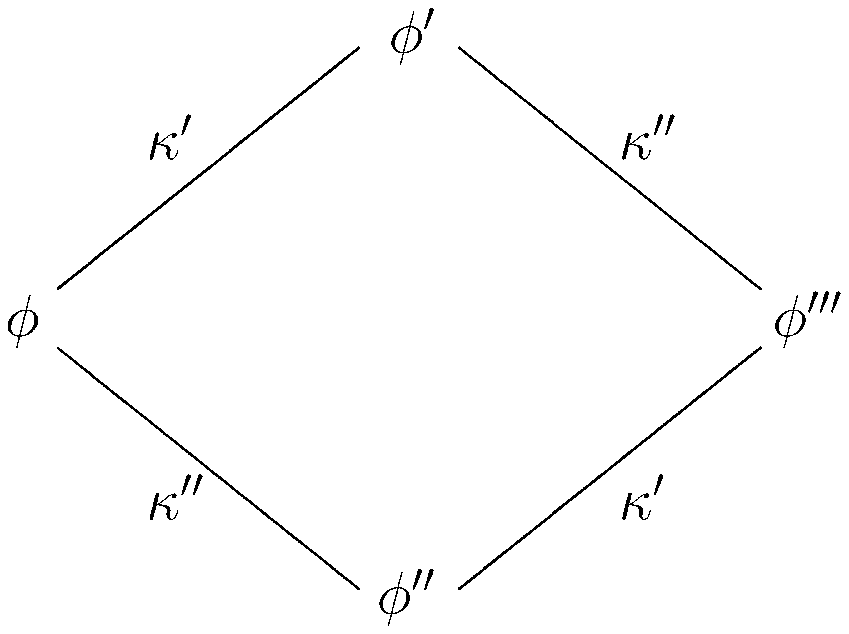,width=9.0cm}  
        \caption{Lamb-Bianchi diagram for four arbitrary solutions $\phi $, $\phi ^{\prime }$, $\phi ^{\prime \prime }$, $\phi ^{\prime \prime
\prime }$ of the sine-Gordon equation, with each link representing a Bäcklund transformation (2.1) with constant $\kappa$ chosen as indicated.}
        \label{Lamb}}
\end{center}

Eliminating all differentials leads to the \emph{nonlinear superposition
principle}\ for the sine-Gordon equation%
\begin{equation}
\phi ^{\prime \prime \prime }=4\arctan \left[ \frac{\kappa ^{\prime }+\kappa
^{\prime \prime }}{\kappa ^{\prime }-\kappa ^{\prime \prime }}\tan \left( 
\frac{\phi ^{\prime }-\phi ^{\prime \prime }}{4}\right) \right] +\phi .
\label{NLS}
\end{equation}%
Hence a new solution $\phi ^{\prime \prime \prime }$ can be constructed in
the functional way as described by (\ref{NLS}) from three known solutions $%
\phi $, $\phi ^{\prime }$, $\phi ^{\prime \prime }$. Moreover, one can
easily see that one can not directly obtain degenerate solutions from (\ref%
{NLS}) as that would require $\phi ^{\prime }=\phi ^{\prime \prime }$ and $%
\kappa ^{\prime }=\kappa ^{\prime \prime }$. We will now demonstrate how the
limits may be taken appropriately, thus leading to degenerate multi-soliton
solutions. Subsequently we study the properties of this new type of
solutions.

At first we construct an $n$-soliton solution with an $(n-1)$-fold
degeneracy. For this purpose we start by relating four solutions to the
sine-Gordon equation as depicted in the Lamb-Bianchi in figure 1 with the
choice $\phi =\phi _{(n-2)\alpha }$, $\phi ^{\prime }=\phi _{(n-1)\alpha }$, 
$\phi ^{\prime \prime }=\phi _{(n-2)\alpha \beta }$ and constants $\kappa
^{\prime }=1/\alpha $ and $\kappa ^{\prime \prime }=1/\beta $, such that by (%
\ref{NLS}) we obtain 
\begin{equation}
\phi _{(n-1)\alpha \beta }=4\arctan \left[ \frac{\alpha +\beta }{\alpha
-\beta }\tan \left( \frac{\phi _{(n-2)\alpha \beta }-\phi _{(n-1)\alpha }}{4}%
\right) \right] +\phi _{(n-2)\alpha }.  \label{super}
\end{equation}%
Using the identity (see appendix for a derivation)%
\begin{equation}
\lim_{\beta \rightarrow \alpha }\frac{\alpha +\beta }{\alpha -\beta }\tan
\left( \frac{\phi _{(n-2)\alpha \beta }-\phi _{(n-1)\alpha }}{4}\right) =%
\frac{\alpha }{2(1-n)}\frac{d\phi _{(n-1)\alpha }}{d\alpha },  \label{Id1}
\end{equation}%
we can perform the non-trivial limit $\beta \rightarrow \alpha $ in (\ref%
{super}), obtaining in this way the recursive equation 
\begin{equation}
\phi _{n\alpha }=\phi _{(n-2)\alpha }-4\arctan \left( \frac{\alpha }{2(n-1)}%
\frac{d\phi _{(n-1)\alpha }}{d\alpha }\right) ,\text{\qquad for }n\geq 2,
\label{rec}
\end{equation}%
with $\phi _{n\alpha }=0$ for $n\leq 0$. In principle equation (\ref{rec})
is sufficient to the degenerate solutions $\phi _{n\alpha }$ recursively.
However, it still involves a derivative term which evidently becomes more
and more complex for higher order. We eliminate this term next and replace
it with combinations of just degenerate solutions. By iterating the B\"{a}%
cklund transformation (\ref{BT}) we compute the derivatives with respect to $%
x$ and $t$ to%
\begin{eqnarray}
\left( \phi _{n\alpha }\right) _{x} &=&2\alpha
\sum\nolimits_{k=1}^{n}(-1)^{n+k}\sin \left( \frac{\phi _{k\alpha }-\phi
_{(k-1)\alpha }}{2}\right) , \\
\left( \phi _{n\alpha }\right) _{t} &=&\frac{2}{\alpha }\sum%
\nolimits_{k=1}^{n}\sin \left( \frac{\phi _{k\alpha }+\phi _{(k-1)\alpha }}{2%
}\right) .
\end{eqnarray}%
Using the relation (see appendix for a derivation) 
\begin{equation}
\alpha \left( \phi _{n\alpha }\right) _{\alpha }=x\left( \phi _{n\alpha
}\right) _{x}-t\left( \phi _{n\alpha }\right) _{t},  \label{axt}
\end{equation}%
we convert this into the derivative with respect to $\alpha $ required in
the recursive relation (\ref{rec})%
\begin{equation}
\frac{\alpha }{2}\frac{d\phi _{(n-1)\alpha }}{d\alpha }=\sum\limits_{k=1}^{n}%
\left[ (-1)^{n+k}x\alpha \sin \left( \frac{\phi _{k\alpha }-\phi
_{(k-1)\alpha }}{2}\right) -\frac{t}{\alpha }\sin \left( \frac{\phi
_{k\alpha }+\phi _{(k-1)\alpha }}{2}\right) \right] .
\end{equation}%
Therefore for $n\geq 2$ equation (\ref{rec}) becomes%
\begin{equation}
\begin{array}{c}
\phi _{n\alpha }=\phi _{(n-2)\alpha }-4\arctan \left[ \frac{1}{1-n}%
\sum\limits_{k=1}^{n-1}\left[ (-1)^{n+k}x\alpha \sin \left( \frac{\phi
_{k\alpha }-\phi _{(k-1)\alpha }}{2}\right) +\frac{t}{\alpha }\sin \left( 
\frac{\phi _{k\alpha }+\phi _{(k-1)\alpha }}{2}\right) \right] \right] .%
\end{array}
\label{tx}
\end{equation}

This equation can be solved iteratively with an appropriate choice for the
initial condition $\phi _{\alpha }$. Taking this to be the well-known kink
solution%
\begin{equation}
\phi _{\alpha }=4\arctan \left( e^{\xi _{+}}\right) \text{,}\qquad \text{%
with }\xi _{\pm }:=t/\alpha \pm x\alpha ,
\end{equation}%
we compute from (\ref{rec}) the degenerate multi-soliton solutions%
\begin{eqnarray}
\phi _{\alpha \alpha } &=&4\arctan \left( \frac{\xi _{-}}{\cosh \xi _{+}}%
\right) ,  \label{deg2} \\
\phi _{\alpha \alpha \alpha } &=&4\arctan \left( \frac{\xi _{+}\cosh \xi
_{+}-\xi _{-}^{2}\sinh \xi _{+}}{\xi _{-}^{2}+\cosh ^{2}\xi _{+}}\right)
+\phi _{\alpha },  \label{deg3} \\
\phi _{\alpha \alpha \alpha \alpha } &=&4\arctan \left[ \frac{-\xi _{-}}{%
3\cosh \xi _{+}}\ \frac{\xi _{-}^{4}+3\xi _{+}^{2}-(3+2\xi _{-}^{2})\cosh
^{2}\xi _{+}+3\xi _{+}\sinh 2\xi _{+}}{\xi _{-}^{4}+\xi _{+}^{2}+2\xi
_{-}^{2}+\cosh ^{2}\xi _{+}-2\xi _{-}^{2}\xi _{+}\tanh \xi _{+}}\right]
+\phi _{\alpha \alpha }~.~~~~~~~~  \label{deg4}
\end{eqnarray}%
We notice that unlike standard soliton solutions these solutions are
governed by different timescales, being exponential and polynomial.
Snapshots of these solutions at two specific values in time are depicted in
figure 2 for some concrete values of $\alpha $. The $N$-kink solution with $%
N=2n$ exhibits $n$ almost identical solitons traveling at nearly the same
speed at small time scales. When $N=2n+1$ the solutions do not vanish
asymptotically and have an additional kink at large values of $x$.

\FIGURE{ \epsfig{file=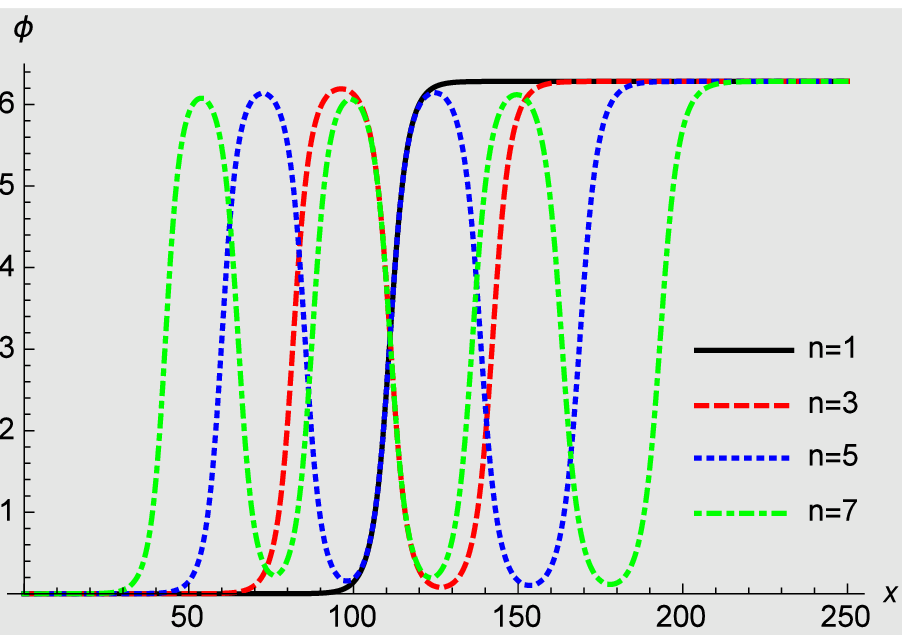,width=7.25cm} \epsfig{file=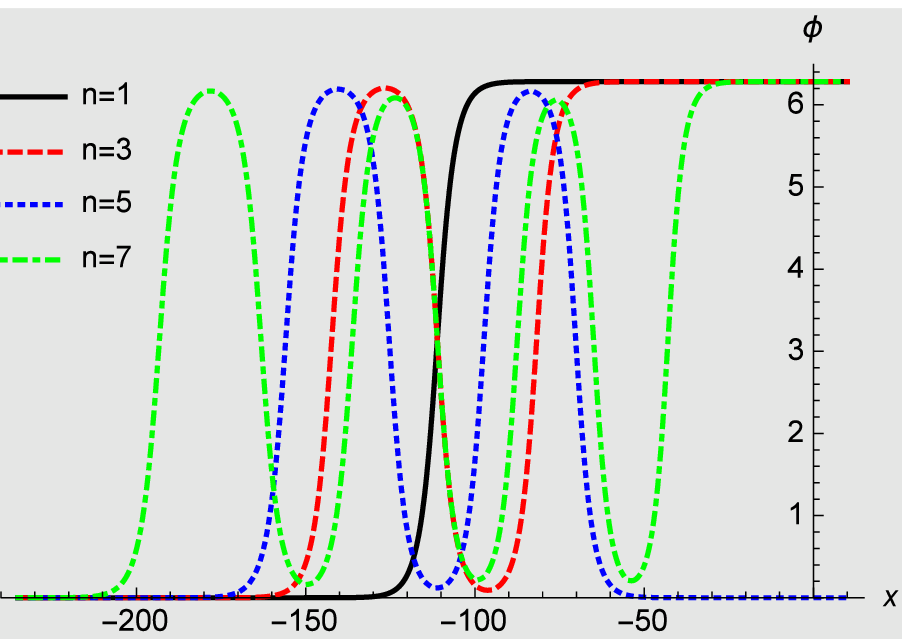,width=7.25cm}
\epsfig{file=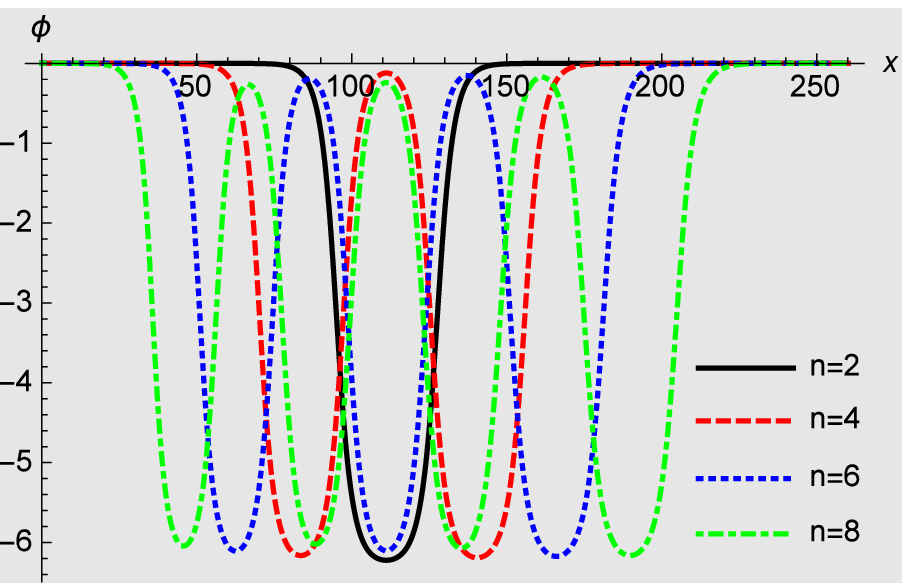,width=7.25cm} \epsfig{file=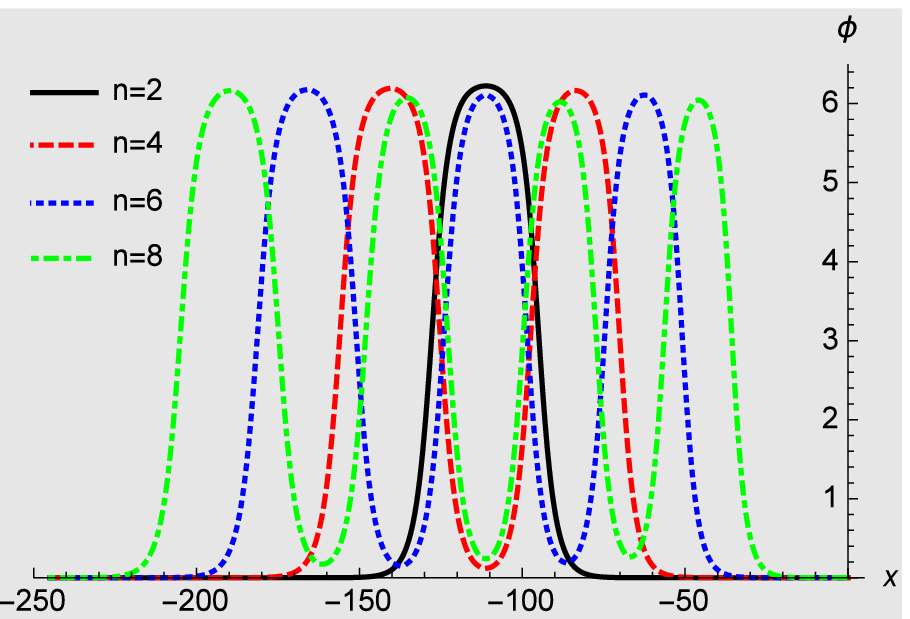,width=7.25cm}
\caption{Sine-Gordon degenerate n-kink solutions at different times $t=-10$ (panel a,c) and $t=10$ (panel b,d) for spectral parameter $\alpha=0.3$.}
        \label{deg16}}

It is clear that solutions constructed in this manner, i.e. by iterating (%
\ref{tx}), will be of a form involving sums over $\arctan $-functions, which
does not immediately allow to study properties such as the asymptotic
behaviour we are interested in here. Of course one may combine these
functions into one using standard identities, which become however
increasingly nested for larger $n$ in $\phi _{n\alpha }$. This can be
avoided by deriving a recursive relation directly for the argument of just
one $\arctan $-function. Defining for this purpose the functions $\tau _{n}$
via the relation%
\begin{equation}
\phi _{n\alpha }=4\arctan \tau _{n},  \label{ft}
\end{equation}%
we convert the recursive relation (\ref{rec}) in $\phi $ into a recursive
relation in $\tau $ as%
\begin{equation}
\tau _{n}=\frac{\tau _{n-2}(1+\tau _{n-1}^{2})-\frac{2\alpha }{n-1}\frac{%
d\tau _{n-1}}{d\alpha }}{1+\tau _{n-1}^{2}+\frac{2\alpha }{n-1}\tau _{n-2}%
\frac{d\tau _{n-1}}{d\alpha }}.  \label{taurec}
\end{equation}%
Similarly as for the derivatives with respect to $\phi $ we may also compute
them for the functions $\tau $ using (\ref{ft}). Computing first 
\begin{eqnarray}
\left( \tau _{n}\right) _{x} &=&\alpha (1+\tau
_{n}^{2})\sum\nolimits_{k=1}^{n}(-1)^{n+k}\frac{(\tau _{k}-\tau
_{k-1})(1+\tau _{k-1}\tau _{k})}{(1+\tau _{k-1}^{2})(1+\tau _{k}^{2})}, \\
\left( \tau _{n}\right) _{t} &=&\frac{1}{\alpha }(1+\tau
_{n}^{2})\sum\nolimits_{k=1}^{n}\frac{(\tau _{k}+\tau _{k-1})(1-\tau
_{k-1}\tau _{k})}{(1+\tau _{k-1}^{2})(1+\tau _{k}^{2})},
\end{eqnarray}%
and using $\alpha \left( \tau _{n}\right) _{\alpha }=x\left( \tau
_{n}\right) _{x}-t\left( \tau _{n}\right) _{t}$ we obtain the derivative of $%
\tau $ with respect to $\alpha $%
\begin{equation}
\frac{\alpha }{(1+\tau _{n}^{2})}\frac{d\tau _{n}}{d\alpha }%
=\sum\limits_{k=1}^{n}\left[ (-1)^{n+k}x\alpha \frac{(\tau _{k}-\tau
_{k-1})(1+\tau _{k-1}\tau _{k})}{(1+\tau _{k-1}^{2})(1+\tau _{k}^{2})}-\frac{%
t}{\alpha }\frac{(\tau _{k}+\tau _{k-1})(1-\tau _{k-1}\tau _{k})}{(1+\tau
_{k-1}^{2})(1+\tau _{k}^{2})}\right] ,
\end{equation}%
which we use to convert (\ref{taurec}) into%
\begin{equation}
\tau _{n}=\frac{\tau _{n-2}-\frac{2}{n-1}\sum\limits_{k=1}^{n-1}\left[
(-1)^{n+k-1}x\alpha \frac{(\tau _{k}-\tau _{k-1})(1+\tau _{k-1}\tau _{k})}{%
(1+\tau _{k-1}^{2})(1+\tau _{k}^{2})}-\frac{t}{\alpha }\frac{(\tau _{k}+\tau
_{k-1})(1-\tau _{k-1}\tau _{k})}{(1+\tau _{k-1}^{2})(1+\tau _{k}^{2})}\right]
}{1+\frac{2\tau _{n-2}}{n-1}\sum\limits_{k=1}^{n-1}\left[ (-1)^{n+k-1}x%
\alpha \frac{(\tau _{k}-\tau _{k-1})(1+\tau _{k-1}\tau _{k})}{(1+\tau
_{k-1}^{2})(1+\tau _{k}^{2})}-\frac{t}{\alpha }\frac{(\tau _{k}+\tau
_{k-1})(1-\tau _{k-1}\tau _{k})}{(1+\tau _{k-1}^{2})(1+\tau _{k}^{2})}\right]
}.
\end{equation}%
Using the variables $\xi _{\pm }$ instead of $x$, $t$ we obtain 
\begin{equation}
\tau _{n}=\frac{\tau _{n-2}-\frac{2}{n-1}\sum\limits_{k=1}^{n-1}\left[ \frac{%
\tau _{k}(\tau _{k-1}^{2}-1)\xi _{(-1)^{n+k}}+\tau _{k-1}(\tau
_{k}^{2}-1)\xi _{(-1)^{n+k+1}}}{(1+\tau _{k-1}^{2})(1+\tau _{k}^{2})}\right] 
}{1+\frac{2\tau _{n-2}}{n-1}\sum\limits_{k=1}^{n-1}\left[ \frac{\tau
_{k}(\tau _{k-1}^{2}-1)\xi _{(-1)^{n+k}}+\tau _{k-1}(\tau _{k}^{2}-1)\xi
_{(-1)^{n+k+1}}}{(1+\tau _{k-1}^{2})(1+\tau _{k}^{2})}\right] }.
\label{tauc}
\end{equation}

These relations lead to simpler compact expressions allowing us to study the
asymptotic properties of these functions more easily. Iterating (\ref{taurec}%
) we obtain the first solutions as%
\begin{eqnarray}
\tau _{1} &=&e^{\xi _{+}},  \label{tau1} \\
\tau _{2} &=&\frac{2\xi _{-}\tau _{1}}{1+\tau _{1}^{2}}=\frac{\xi _{-}}{%
\cosh \xi _{+}}, \\
\tau _{3} &=&\frac{(1+2\xi _{+}+2\xi _{-}^{2})\tau _{1}+\tau _{1}^{3}}{%
1+(1-2\xi _{+}+2\xi _{-}^{2})\tau _{1}^{2}}, \\
\tau _{4} &=&\frac{4\xi _{-}(3+3\xi _{+}+\xi _{-}^{2})\tau _{1}+4\xi
_{-}(3-3\xi _{+}+\xi _{-}^{2})\tau _{1}^{3}}{3+(6+12\xi _{+}^{2}+4\xi
_{-}^{4})\tau _{1}^{2}+3\tau _{1}^{4}}, \\
\tau _{5} &=&\frac{3c_{+}\tau _{1}+2d_{+}\tau _{1}^{3}+9\tau _{1}^{5}}{%
9+2\tau _{1}^{2}d_{-}+3c_{-}\tau _{1}^{4}},  \label{tau5}
\end{eqnarray}%
with%
\begin{eqnarray*}
c_{\pm } &=&6\xi _{+}^{2}\pm 12\xi _{+}\xi _{-}^{2}\pm 12\xi _{+}+2\xi
_{-}^{4}+18\xi _{-}^{2}+3, \\
d_{\pm } &=&\pm 18\xi _{+}^{3}+18\xi _{+}^{2}\xi _{-}^{2}-9\xi _{+}^{2}\mp
6\xi _{+}\xi _{-}^{4}\mp 36\xi _{+}\xi _{-}^{2}\pm 18\xi _{+}+2\xi
_{-}^{6}+3\xi _{-}^{4}+27\xi _{-}^{2}+9.
\end{eqnarray*}%
It is now straightforward to compute the $\tau _{n}$ for any larger value of 
$n$ in this manner.

It is clear that by setting up the nonlinear superposition equation (\ref%
{super}) for different types of solutions will produce recurrence relations
for new types of degenerate multi-soliton solutions. We will not pursue this
here, but instead compare the results obtained in this section with those
obtained from different type of methods.

\section{Degenerate multi-solitons from Darboux-Crum transformations}

Classical integrability is build into the zero curvature condition, or
AKNS-equation \cite{AKNS}, for two operators $U$ and $V$, which can be
expressed equivalently in terms of two linear first order differential
equations for an auxiliary function $\Psi $%
\begin{equation}
\partial _{t}U-\partial _{x}V+\left[ U,V\right] =0\qquad \Leftrightarrow
\qquad \Psi _{t}=V\Psi \text{, }\Psi _{x}=U\Psi .  \label{ZC}
\end{equation}%
When \cite{AKNSsg} taking the matrix valued functions $U$, $V$ and the
column vector $\Psi $ to be of the \ form%
\begin{equation}
U=\left( 
\begin{array}{cc}
\frac{i}{2}\phi _{x} & \frac{\alpha }{2} \\ 
\frac{\alpha }{2} & -\frac{i}{2}\phi _{x}%
\end{array}%
\right) ,\qquad V=\left( 
\begin{array}{cc}
0 & \frac{1}{2\alpha }e^{i\phi } \\ 
\frac{1}{2\alpha }e^{-i\phi } & 0%
\end{array}%
\right) ,\qquad \Psi =\left( 
\begin{array}{c}
\psi \\ 
\varphi%
\end{array}%
\right) ,
\end{equation}%
the zero curvature condition (\ref{ZC}) holds when the field $\phi $
satisfies the sine-Gordon equation (\ref{SG}). Thus we have the four
auxiliary equations%
\begin{equation}
\psi _{x}=\frac{i}{2}\phi _{x}\psi +\frac{\alpha }{2}\varphi ,\quad \varphi
_{x}=\frac{\alpha }{2}\psi -\frac{i}{2}\phi _{x}\varphi ,\quad \psi _{t}=%
\frac{1}{2\alpha }e^{i\phi }\varphi ,\quad \varphi _{t}=\frac{1}{2\alpha }%
e^{-i\phi }\psi .  \label{ZCC}
\end{equation}%
Computing $\psi _{xx}$ and $\varphi _{xx}$ decouples the first order
equations (\ref{ZCC}) into two Schr\"{o}dinger equations 
\begin{equation}
-\psi _{xx}+V_{+}\psi =-\frac{\alpha ^{2}}{4}\psi ,\qquad -\varphi
_{xx}+V_{-}\varphi =-\frac{\alpha ^{2}}{4}\varphi  \label{sch}
\end{equation}%
with potentials%
\begin{equation}
V_{\pm }=-\frac{1}{4}(\phi _{x})^{2}\pm \frac{i}{2}\phi _{xx}.  \label{pot}
\end{equation}%
We notice that for real sine-Gordon fields $\phi $ the potentials are
complex, whereas real potentials may be obtained from purely imaginary
fields $\phi $. However, even for complex potentials the invariance of these
equations under the antilinear involution maps, or $\mathcal{PT}$-symmetries,%
\begin{equation}
\mathcal{PT}_{\mu }:x\rightarrow -x\text{, }t\rightarrow -t\text{, }%
i\rightarrow -i\text{, }\phi \rightarrow -\phi \text{, }\chi \rightarrow
e^{i\mu }\chi \text{, }~~~~~~\text{for }\chi =\psi ,\varphi ,  \label{PT}
\end{equation}%
will guarantee the spectral parameter $\alpha $ to be real \cite%
{EW,Bender:1998ke,Rev2,Alirev}. Note that $\mu $ might be different for the
two equations in (\ref{sch}).

Taking now $\psi _{\alpha _{0}}$, $\varphi _{\alpha _{0}}$ and $\psi
_{\alpha _{1}}$, $\varphi _{\alpha _{1}}$ to be solutions of (\ref{ZCC})
with spectral parameters $\alpha _{0}$ and $\alpha _{1}$, respectively, but
the same sine-Gordon field $\phi =\phi ^{(0)}$, the Darboux transformations 
\cite{darboux,matveevdarboux} yield the new solutions for the two Schr\"{o}%
dinger equations (\ref{sch}) 
\begin{equation}
\psi _{\alpha _{0},\alpha _{1}}^{(1)}=\alpha _{0}\varphi _{\alpha
_{0}}-\alpha _{1}\frac{\varphi _{\alpha _{1}}}{\psi _{\alpha _{1}}}\psi
_{\alpha _{0}},\quad \text{and\quad }\varphi _{\alpha _{0},\alpha
_{1}}^{(1)}=\alpha _{0}\psi _{\alpha _{0}}-\alpha _{1}\frac{\psi _{\alpha
_{1}}}{\varphi _{\alpha _{1}}}\varphi _{\alpha _{0}},
\end{equation}%
accompanied by the new solution for the sine-Gordon equation%
\begin{equation}
\phi _{\alpha _{0},\alpha _{1}}^{(1)}=\phi ^{(0)}-2i\ln \frac{\varphi
_{\alpha _{1}}}{\psi _{\alpha _{1}}},  \label{level1}
\end{equation}%
involving two spectral parameters.

The iteration of this procedure is usually referred to as the Darboux--Crum
transformation \cite{crum,matveevdarboux} leading to the new solutions%
\begin{equation}
\psi _{\alpha _{0},\alpha _{1},\ldots ,\alpha _{n}}^{(n)}=\frac{W\left[ \psi
_{\alpha _{0}},\psi _{\alpha _{1}},\ldots ,\psi _{\alpha _{n}}\right] }{W%
\left[ \psi _{\alpha _{1}},\ldots ,\psi _{\alpha _{n}}\right] },\qquad
\varphi _{\alpha _{0},\alpha _{1},\ldots ,\alpha _{n}}^{(n)}=\frac{W\left[
\varphi _{\alpha _{0}},\varphi _{\alpha _{1}},\ldots ,\varphi _{\alpha _{n}}%
\right] }{W\left[ \varphi _{\alpha _{1}},\ldots ,\varphi _{\alpha _{n}}%
\right] },
\end{equation}%
together with%
\begin{equation}
\phi _{\alpha _{0},\alpha _{1},\ldots ,\alpha _{n}}^{(n)}=\phi ^{(0)}-2i\ln 
\frac{W\left[ \varphi _{\alpha _{0}},\varphi _{\alpha _{1}},\ldots ,\varphi
_{\alpha _{n}}\right] }{W\left[ \psi _{\alpha _{1}},\ldots ,\psi _{\alpha
_{n}}\right] }.  \label{DC}
\end{equation}%
Here $W[\cdot ]$ denotes the Wronskians, e.g. $W[f,g]=fg_{x}-gf_{x}$ or in
general $W[f_{1},f_{2},\ldots ,f_{n}]=\det w$ with $%
w_{ij}=d^{i-1}f_{j}/dx^{i-1}$ for $i,j=1,\ldots ,n$.

Next we carry out the limit to the degenerate case by using Jordan states in
the Wronskian similarly as discussed in more detail for the KdV-equation in 
\cite{CorreaFring}. In general \emph{Jordan states} $\Xi _{\lambda }^{(k)}$
are defined as the solutions of the iterated Schr\"{o}dinger equation 
\begin{equation}
\hat{H}^{k+1}\Xi _{\lambda }^{(k)}=\left[ -\partial _{x}^{2}+V-E(\lambda )%
\right] ^{k+1}\Xi _{\lambda }^{(k)}=0,  \label{Jordan}
\end{equation}%
with eigenvalue $E(\lambda )$ depending on the spectral parameter $\lambda $
and potential $V$. This means for $k=0$ the Jordan states are just the
eigenfunction of the Schr\"{o}dinger equation, namely $\Xi _{\lambda
}^{(0)}=\psi _{\lambda }~$or $\Xi _{\lambda }^{(0)}=\phi _{\lambda }$ with $%
\phi _{\lambda }$ being the second fundamental solution to the same
eigenvalue $E(\lambda )$. It can be constructed from Liouville's formula $%
\phi _{\lambda }(x)=\psi _{\lambda }(x)\dint\nolimits^{x}\left[ \psi
_{\lambda }(s)\right] ^{-2}ds$ from the first solution $\psi _{\lambda }$.
The general solution to (\ref{Jordan}) is%
\begin{equation}
\Xi _{\lambda }^{(k)}=\dsum\limits_{l=0}^{k}c_{l}\chi _{\lambda
}^{(l)}+\dsum\limits_{l=0}^{k}d_{l}\Omega _{\lambda }^{(l)},\qquad
c_{l},d_{l}\in \mathbb{R},  \label{sol}
\end{equation}%
with $\chi _{\lambda }^{(k)}:=\partial ^{k}\psi _{\lambda }/\partial E^{k}$
and $\Omega _{\lambda }^{(k)}:=\partial ^{k}\phi _{\lambda }/\partial E^{k}$.

\subsection{Kinks, antikinks, breather and imaginary cusps from vanishing
potentials}

We start by solving the four linear first order differential equations (\ref%
{ZCC}) to the lowest level in the Darboux-Crum iteration procedure for some
specific choices of $\phi ^{(0)}$. Considering the simplest case of
vanishing potentials $V_{\pm }=0$, by taking $\phi ^{(0)}=2\pi \lambda $
with $\lambda \in \mathbb{Z}$, the equations in (\ref{ZCC}) are easily
solved by%
\begin{equation}
\psi _{\alpha }(x,t)=c_{1}e^{\xi _{+}/2}+c_{2}e^{-\xi _{+}/2}\quad \text{%
and\quad }\varphi _{\alpha }(x,t)=c_{1}e^{\xi _{+}/2}-c_{2}e^{-\xi _{+}/2}.
\label{psiphi}
\end{equation}%
Evidently, the constants $c_{1}$, $c_{2}\in \mathbb{C}$ are taking care
about the boundary conditions. Imposing the $\mathcal{PT}_{\mu }$-symmetry
as defined in (\ref{PT}) on each of these solutions selects out some
specific choices for the constants. For instance, for $\lambda =0$ and $%
c_{1}=c_{2}$ the fields in (\ref{psiphi}) obey the symmetries $\mathcal{PT}%
_{0}:\psi _{\alpha }\rightarrow \psi _{\alpha }$, $\mathcal{PT}_{\pi
}:\varphi _{\alpha }\rightarrow -\varphi _{\alpha }$ and we obtain from the
Darboux transformation (\ref{level1}) the purely imaginary cusp solution 
\begin{equation}
\phi _{\alpha }^{c}=-2i\ln \left( \tanh \left\vert \xi _{+}\right\vert
/2\right) .  \label{s1}
\end{equation}%
Imposing instead the symmetries $\mathcal{PT}_{-\pi /2}:\psi _{\alpha
}\rightarrow i\psi _{\alpha }$, $\mathcal{PT}_{-\pi /2}:\varphi _{\alpha
}\rightarrow -i\varphi _{\alpha }~$on the fields in (\ref{psiphi}) the kink
and antikink solutions 
\begin{equation}
\phi _{\alpha }=4\arctan \left( e^{\xi _{+}}\right) \qquad \text{and\qquad }%
\phi _{\bar{\alpha}}=4\func{arccot}\left( e^{\xi _{+}}\right) ,  \label{s2}
\end{equation}%
are obtained from $\lambda =1$ and $c_{2}=ic_{1}\in \mathbb{R}$ and $\lambda
=0$ and $c_{1}=ic_{2}\in \mathbb{R}$, respectively. We notice that the
remaining constant $c_{1}$ or $c_{2}$ cancel out in all solutions in (\ref%
{s1}) and (\ref{s2}). Iterating these results leads for instance to

\subsubsection{Degenerate kink solutions}

For the choice $c_{2}=ic_{1}$ we obtain the degenerate solutions 
\begin{equation}
\phi _{na}=-2i\ln \frac{W\left[ \varphi _{\alpha },\partial _{\alpha
}\varphi _{\alpha },\partial _{\alpha }^{2}\varphi _{\alpha },\ldots
,\partial _{\alpha }^{n-1}\varphi _{\alpha }\right] }{W\left[ \psi _{\alpha
},\partial _{\alpha }\psi _{\alpha },\partial _{\alpha }^{2}\psi _{\alpha
},\ldots ,\partial _{\alpha }^{n-1}\psi _{\alpha }\right] },
\end{equation}%
which when evaluated explicitly coincide precisely with the expressions
previously obtained in (\ref{deg2})-(\ref{deg4}).

\subsubsection{Degenerate complex cusp solutions}

In a similar way we can construct degenerate purely complex cusp solutions.
Such type of solutions are also well-known in the literature, see for
instance \cite{kawamoto} for an early occurrence. These solutions appear to
be non-physical at first sight, but they find applications for instance as
an explanation for the entrainment of air \cite{eggers2001air}. For the
choice $c_{1}=c_{2}$ we obtain%
\begin{eqnarray}
\phi _{2a}^{c} &=&-2i\ln \frac{W\left[ \varphi _{\alpha },\partial _{\alpha
}\varphi _{\alpha }\right] }{W\left[ \psi _{\alpha },\partial _{\alpha }\psi
_{\alpha }\right] }=-2i\ln \left( \frac{\sinh \xi _{+}+\xi _{-}}{\sinh \xi
_{+}-\xi _{-}}\right) , \\
\phi _{3a}^{c} &=&-2i\ln \frac{W\left[ \varphi _{\alpha },\partial _{\alpha
}\varphi _{\alpha },\partial _{\alpha }^{2}\varphi _{\alpha }\right] }{W%
\left[ \psi _{\alpha },\partial _{\alpha }\psi _{\alpha },\partial _{\alpha
}^{2}\psi _{\alpha }\right] } \\
&=&-2i\ln \left( \frac{\cosh \left( 3\xi _{+}/2\right) +2\xi _{+}\sinh
\left( \xi _{+}/2\right) -(1+2\xi _{-}^{2})\cosh \left( \xi _{+}/2\right) }{%
\sinh \left( 3\xi _{+}/2\right) -2\xi _{+}\cosh \left( \xi _{+}/2\right)
+(1+2\xi _{-}^{2})\sinh \left( \xi _{+}/2\right) }\right) .  \notag
\end{eqnarray}%
Similarly we can proceed to obtain the solutions $\phi _{na}^{c}$ for $n>3$.

\subsubsection{Degenerate breathers}

Breather solutions may be obtained in various ways. An elegant real solution
can be constructed as follows: Taking as the starting point the two-kink
solution with two distinct spectral parameters $\alpha $ and $\beta $%
\begin{equation}
\phi _{\alpha \beta }=-2i\ln \frac{W\left[ \varphi _{\alpha },\varphi
_{\beta }\right] }{W\left[ \psi _{\alpha },\psi _{\beta }\right] }=4\arctan %
\left[ \frac{\alpha +\beta }{\alpha -\beta }\frac{\sinh \left[ (\frac{1}{%
2\beta }-\frac{1}{2\alpha })(t-x\alpha \beta )\right] }{\cosh \left[ (\frac{1%
}{2\beta }+\frac{1}{2\alpha })(t+x\alpha \beta )\right] }\right] ,
\end{equation}%
we obtain a breather by converting one of the functions in the argument into
a trigonometric function. Taking first $\beta \rightarrow 1/\alpha $ we
obtain%
\begin{equation}
\phi _{\alpha ,1/\alpha }=4\arctan \left[ \frac{\alpha ^{2}+1}{\alpha ^{2}-1}%
\frac{\sinh \left[ \frac{1}{2}(\alpha -\frac{1}{\alpha })(t-x)\right] }{%
\cosh \left[ \frac{1}{2}(\alpha +\frac{1}{\alpha })(t+x)\right] }\right] .
\end{equation}%
Thus by demanding that $(\alpha ^{2}+1)/(\alpha ^{2}-1)=i\theta $ and $%
(\alpha -\frac{1}{\alpha })/2=-i\tilde{\theta}$ for some constants for $%
\theta ,\tilde{\theta}\in \mathbb{R}$ we obtain an oscillatory function in
the argument of the $\arctan $. Solving for instance the first relation
gives $\alpha =\tilde{\alpha}=:(\theta -i)/\sqrt{1+\theta ^{2}}$ so that $%
\tilde{\theta}=1/\sqrt{1+\theta ^{2}}$. The corresponding breather solution
then results to%
\begin{equation}
\phi _{\tilde{\alpha},1/\tilde{\alpha}}=4\arctan \left[ \theta \frac{\sin %
\left[ (t-x)/\sqrt{1+\theta ^{2}}\right] }{\cosh \left[ \theta (t+x)/\sqrt{%
1+\theta ^{2}}\right] }\right] .  \label{b1}
\end{equation}%
This solution evolves with a constant speed $-1$ modulated by some overall
oscillation resulting from the $\sin $-function. Similarly we can construct
a two-breather solution from two degenerated kink-solutions 
\begin{equation}
\phi _{\alpha \alpha \beta \beta }=-2i\ln \frac{W\left[ \varphi _{\alpha
},\partial _{\alpha }\varphi _{\alpha },\varphi _{\beta },\partial _{\beta
}\varphi _{\beta }\right] }{W\left[ \psi _{\alpha },\partial _{\alpha }\psi
_{\alpha },\psi _{\beta },\partial _{\beta }\psi _{\beta }\right] },
\end{equation}%
by using the same parameterization $\phi _{\tilde{\alpha},\tilde{\alpha},1/%
\tilde{\alpha},1/\tilde{\alpha}}$.

\subsection{Cnoidal kink solutions from shifted Lam\'{e} potentials}

The sine-Gordon equation also admits a solution in terms of the Jacobi
amplitude $\limfunc{am}(x,m)$ depending on the parameter $0\leq m\leq 1$ in
the form%
\begin{equation}
\phi _{cn}^{(0)}=2\limfunc{am}\left( \frac{x-t}{\sqrt{\mu }},\mu \right)
\label{fcn}
\end{equation}%
for any $0<\mu (m)<1$. Denoting the Jacobi elliptic functions as $\limfunc{cn%
}(x,m)=\cos [\limfunc{am}(x,m)]$, $\limfunc{sn}(x,m)=\sin [\limfunc{am}%
(x,m)] $ and $\limfunc{dn}(x,m)=(1-m\sin ^{2}[\limfunc{am}(x,m)])^{1/2}$
this follows directly from $\limfunc{am}^{\prime }(x,m)=\limfunc{dn}(x,m)$, $%
\limfunc{dn}^{\prime }(x,m)=-m\limfunc{sn}(x,m)\limfunc{cn}(x,m)$ and the
double angle identity for the $\sin $-function. The potentials (\ref{pot})
following from the solution (\ref{fcn}) are 
\begin{eqnarray}
V_{\pm }^{cn} &=&-\frac{1}{\mu }\limfunc{dn}{}^{2}\left( \frac{x-t}{\sqrt{%
\mu }},\mu \right) \mp i\limfunc{cn}{}\left( \frac{x-t}{\sqrt{\mu }},\mu
\right) \limfunc{sn}\left( \frac{x-t}{\sqrt{\mu }},\mu \right) \\
&=&\frac{\sqrt{m}}{2}\limfunc{sn}{}^{2}\left( \frac{x-t}{2m^{1/4}}\mp \frac{i%
}{2}K^{\prime },m\right) -\frac{1}{4}(m^{1/2}+m^{-1/2}),
\end{eqnarray}%
where we used the parameterization $\mu =4\sqrt{m}/(1+\sqrt{m})^{2}$ with $%
K(m)$ denoting the complete elliptic integral of the first kind and $%
K^{\prime }(m)=K(1-m)$. Notice that this is a complex shifted and scaled Lam%
\'{e} potential \cite{ince1940vii,whittaker} invariant under any $\mathcal{PT%
}$-symmetry as defined in (\ref{PT}). Such type of potentials emerge in
various contexts, e.g. they give rise to elliptic string solutions in $%
AdS_{3}~$and $dS_{3}$ \cite{bakas2016elliptic} or the study of the origin of
spectral singularities in periodic $\mathcal{PT}$-symmetric systems \cite%
{correa2012l}.

Next we need to find the solutions $\psi $ and $\varphi $ to the
AKNS-equations (\ref{ZCC}) corresponding to the sine-Gordon solution $\phi
_{cn}^{(0)}$. We need to distinguish the two cases $0\leq \alpha \leq 1$ and 
$\alpha >1$. In the first case we parameterize $\alpha =m^{1/4}$ finding the
solutions%
\begin{equation}
\psi _{m}^{<}(x,t)=c\limfunc{cn}{}\left[ \frac{x-t}{2m^{1/4}}-\frac{i}{2}%
K^{\prime },m\right] ,\quad \varphi _{m}^{<}(x,t)=-ic\limfunc{cn}{}\left[ 
\frac{x-t}{2m^{1/4}}+\frac{i}{2}K^{\prime },m\right]  \label{scon1}
\end{equation}%
and for the second case we parameterize $\alpha =m^{-1/4}$ obtaining the
solutions%
\begin{equation}
\psi _{m}^{>}(x,t)=ic\limfunc{dn}{}\left[ \frac{x-t}{2m^{1/4}}-\frac{i}{2}%
K^{\prime },m\right] ,\quad \varphi _{m}^{>}(x,t)=c\limfunc{dn}{}\left[ 
\frac{x-t}{2m^{1/4}}+\frac{i}{2}K^{\prime },m\right] ,  \label{scon2}
\end{equation}%
with integration constant $c$. For real values of $c$ we observe the $%
\mathcal{PT}$-symmetries $\mathcal{PT}_{0}\psi _{m}^{<}=\psi _{m}^{<}$, $%
\mathcal{PT}_{\pi }\varphi _{m}^{<}=\varphi _{m}^{<}$, $\mathcal{PT}_{\pi
}\psi _{m}^{>}=\psi _{m}^{>}$, $\mathcal{PT}_{0}\varphi _{m}^{>}=\varphi
_{m}^{>}$.

The Darboux transformation (\ref{level1}) then yields the real solutions for
the sine-Gordon equation%
\begin{eqnarray}
\phi _{<,m}^{(1)}(x,t) &=&2\limfunc{am}\left( \frac{x-t}{\sqrt{\mu }},\mu
\right) -4\arctan \left[ \frac{\limfunc{dn}{}\left( \frac{x-t}{2m^{1/4}}%
,m\right) \limfunc{sn}\left( \frac{x-t}{2m^{1/4}},m\right) }{\limfunc{cn}{}%
\left( \frac{x-t}{2m^{1/4}},m\right) }\right] -\pi ,  \label{phs1} \\
\phi _{>,m}^{(1)}(x,t) &=&2\limfunc{am}\left( \frac{x-t}{\sqrt{\mu }},\mu
\right) -4\arctan \left[ \sqrt{m}\frac{\limfunc{cn}{}\left( \frac{x-t}{%
2m^{1/4}},m\right) \limfunc{sn}\left( \frac{x-t}{2m^{1/4}},m\right) }{%
\limfunc{dn}{}\left( \frac{x-t}{2m^{1/4}},m\right) }\right] -\pi ,~~~
\label{phs2}
\end{eqnarray}%
after using the addition theorem for the Jacobi elliptic functions, the
following properties $\limfunc{cn}{}\left( iK^{\prime }/2,m\right) =\sqrt{1+%
\sqrt{m}}/m^{1/4}$, $\limfunc{sn}{}\left( iK^{\prime }/2,m\right) =i/m^{1/4}$%
, $\limfunc{dn}{}\left( iK^{\prime }/2,m\right) =\sqrt{1+\sqrt{m}}$, and the
well known relation between the $\ln $ and the $\arctan $-functions. Notice
that the $\limfunc{cn}$-function can be vanishing for real arguments, such
that $\phi _{<}^{(1)}$ is a discontinuous function. Furthermore, we observe
that this solution has a fixed speed and does not involve any variable
spectral parameter, which is vital for the construction of multi-solutions,
see e.g. \cite{CorreaFring}. For this reason we construct a different type
of solution also related to $\phi _{cn}^{(0)}$ that involves an additional
parameter.

These type of solutions can be obtained from 
\begin{equation}
\Psi _{m,\beta }^{\pm }(x)=\frac{H\left( x\pm \beta \right) }{\Theta (x)}%
e^{\mp xZ(\beta )},\qquad \Phi _{m,\beta }^{\pm }(x)=\frac{\Theta \left(
x\pm \beta \right) }{\Theta (x)}e^{\mp xY(\beta )},  \label{lame}
\end{equation}%
solving the Schr\"{o}dinger equation involving the Lam\'{e} potential $V_{L}$%
\begin{equation}
-\Psi _{xx}+V_{L}\Psi =E_{\beta }\Psi ,~~~\text{with }V_{L}=2m\limfunc{sn}%
\left( x,m\right) ^{2}-(1+m)\text{, }
\end{equation}%
with $E_{\beta }=-m\limfunc{sn}\left( \beta ,m\right) ^{2}$ and $E_{\beta
}=-1/\limfunc{sn}\left( \beta ,m\right) ^{2}$, respectively. The functions $%
H $, $\Theta $, $Y$ and $Z$ are defined in terms of Jacobi's theta functions 
$\vartheta _{i}(z,q)$ with $i=1,2,3,4$, $\kappa =\pi /(2K)$ and nome $q=\exp
(-\pi K/K^{\prime })$ as%
\begin{equation}
H\left( z\right) :=\vartheta _{1}\left( z\kappa ,q\right) ,\quad \Theta
\left( z\right) :=\vartheta _{4}\left( z\kappa ,q\right) ,\quad Y(z):=\kappa 
\frac{H^{\prime }\left( z\kappa \right) }{H\left( z\kappa \right) },\quad
Z(z):=\kappa \frac{\Theta ^{\prime }\left( z\kappa \right) }{\Theta \left(
z\kappa \right) }.
\end{equation}%
With a suitable normalization factor and the introduction of a
time-dependence the function $\Psi ^{\pm }(x)$ can be tuned to solve the
equations (\ref{ZCC}). We find%
\begin{eqnarray}
\Psi _{\pm ,m,\beta }^{<}(x,t) &=&\Psi _{m,\beta }^{\pm }\left( \frac{x-t}{%
2m^{1/4}}-\frac{i}{2}K^{\prime }\right) e^{\pm \frac{t}{2m^{1/4}}\frac{%
\limfunc{cn}\left( \beta ,m\right) \limfunc{dn}\left( \beta ,m\right) }{%
\limfunc{sn}\left( \beta ,m\right) }},  \label{sp1} \\
\Phi _{\pm ,m,\beta }^{<}(x,t) &=&\mp e^{\pm iK^{\prime }Z(\beta )\pm i\beta
\kappa }\Psi _{m,\beta }^{\pm }\left( \frac{x-t}{2m^{1/4}}+\frac{i}{2}%
K^{\prime }\right) e^{\pm \frac{t}{2m^{1/4}}\frac{\limfunc{cn}\left( \beta
,m\right) \limfunc{dn}\left( \beta ,m\right) }{\limfunc{sn}\left( \beta
,m\right) }},  \label{sp2}
\end{eqnarray}%
for $\alpha =m^{1/4}\limfunc{sn}\left( \beta ,m\right) $ and%
\begin{eqnarray}
\Psi _{\pm ,m,\beta }^{>}(x,t) &=&\Phi _{m,\beta }^{\pm }\left( \frac{x-t}{%
2m^{1/4}}-\frac{i}{2}K^{\prime }\right) e^{\mp \frac{t}{2m^{1/4}}\frac{%
\limfunc{cn}\left( \beta ,m\right) \limfunc{dn}\left( \beta ,m\right) }{%
\limfunc{sn}\left( \beta ,m\right) }}  \label{sp3} \\
\Phi _{\pm ,m,\beta }^{>}(x,t) &=&\mp e^{\pm iK^{\prime }Y(\beta )\pm i\beta
\kappa }\Phi _{m,\beta }^{\pm }\left( \frac{x-t}{2m^{1/4}}+\frac{i}{2}%
K^{\prime }\right) e^{\mp \frac{t}{2m^{1/4}}\frac{\limfunc{cn}\left( \beta
,m\right) \limfunc{dn}\left( \beta ,m\right) }{\limfunc{sn}\left( \beta
,m\right) }}  \label{sp4}
\end{eqnarray}%
for $\alpha =1/(m^{1/4}\limfunc{sn}\left( \beta ,m\right) )$. The
corresponding solutions for the sine-Gordon equation resulting from the
Darboux transformation (\ref{level1}) are%
\begin{equation}
\phi _{\pm ,m,\beta }^{\ell (1)}(x,t)=\phi _{cn}^{(0)}\pm 2\beta \kappa
-4\arctan \left[ i\frac{M_{\pm }^{\ell }-(M_{\pm }^{\ell })^{\ast }}{M_{\pm
}^{\ell }+(M_{\pm }^{\ell })^{\ast }}\right] ,~~\ ~\ell =<,>
\end{equation}%
with the abbreviations%
\begin{eqnarray}
M_{\pm }^{<} &=&H\left( \frac{x-t}{2m^{1/4}}\pm \beta +\frac{i}{2}K^{\prime
}\right) \Theta \left( \frac{x-t}{2m^{1/4}}-\frac{i}{2}K^{\prime }\right) ,
\\
M_{\pm }^{>} &=&\Theta \left( \frac{x-t}{2m^{1/4}}\pm \beta +\frac{i}{2}%
K^{\prime }\right) \Theta \left( \frac{x-t}{2m^{1/4}}-\frac{i}{2}K^{\prime
}\right) .
\end{eqnarray}

\noindent We depict this solution in figure \ref{degcniodal}. We notice that
the two solutions depicted are qualitatively very similar and appear to be
just translated in amplitude and $x$. However, these translations are not
exact and even the approximations depend nontrivially on $\beta $ and $m$.

\FIGURE{ \epsfig{file=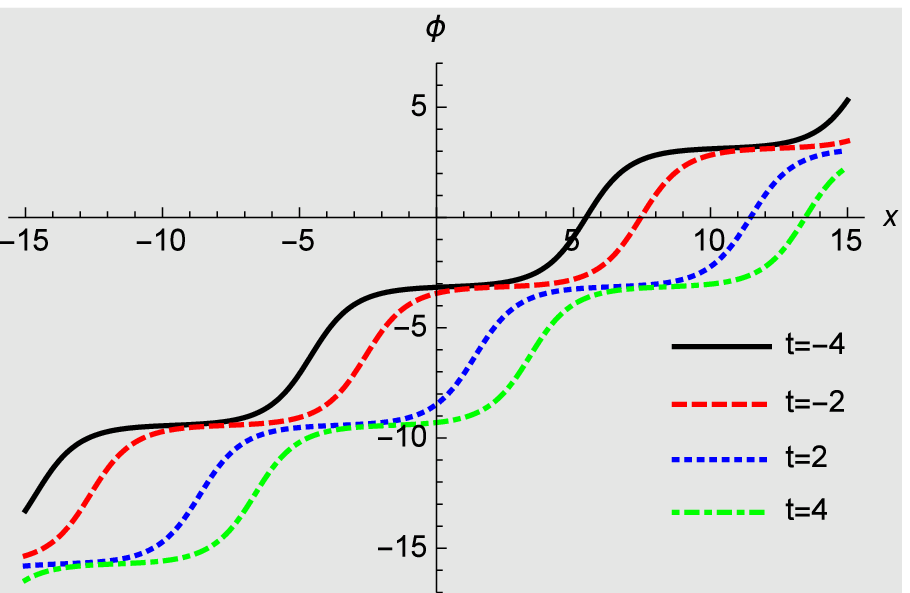,width=7.2cm} \epsfig{file=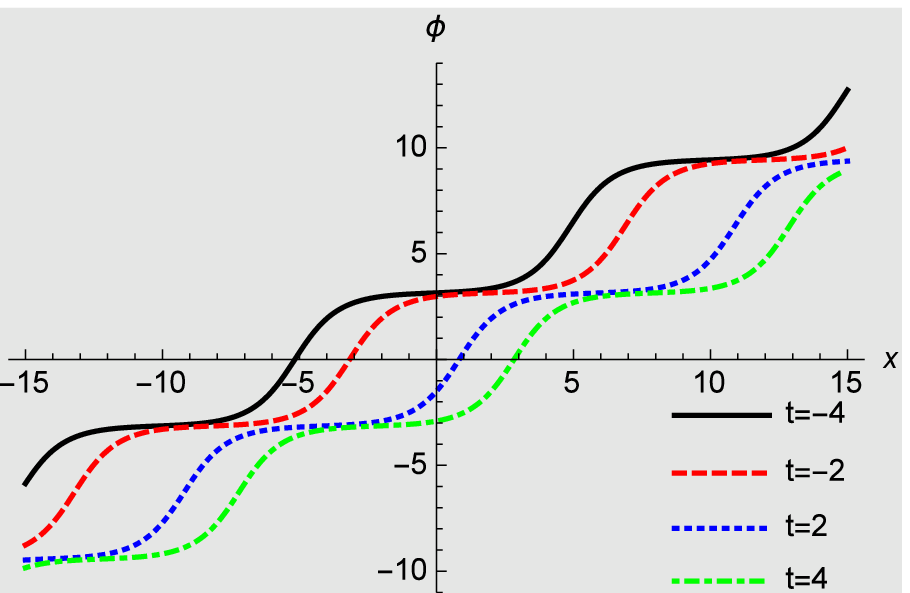,width=7.2cm}
\caption{Sine-Gordon cnoidal kink solution $\phi _{+ ,m,\beta }^{> (1)}(x,t)$ (panel a) and degenerate cnoidal kink solution $\phi _{+ ,m,\beta \beta }^{c>}(x,t)$ (panel b) for spectral parameter $\beta=0.9$ and $m=0.3$ at different times.}
        \label{degcniodal}}

Taking the normalization constants in (\ref{scon1}) and (\ref{scon2})
respectively as $c=\pm m^{1/4}/(1-m)^{1/4}$ and $c=i/(1-m)^{1/4}$ we recover
the simpler solution with constant speed parameter from the limits%
\begin{equation*}
\lim_{\beta \rightarrow K}\Psi _{\pm ,m,\beta }^{\ell }(x,t)=\psi _{m}^{\ell
}(x,t),~\lim_{\beta \rightarrow K}\Phi _{\pm ,m,\beta }^{\ell }(x,t)=\varphi
_{m}^{\ell }(x,t),~\lim_{\beta \rightarrow K}\phi _{\pm ,m,\beta }^{\ell
}(x,t)=\phi _{m}^{\ell }(x,t),~~
\end{equation*}%
such that (\ref{sp1}) and (\ref{sp4}) can be viewed as generalizations of
those solutions.

It is interesting to compare these type of solutions and investigate whether
they can be used to obtain B\"{a}cklund transformations. It is clear that
since $\phi _{cn}^{(0)}$ does not contain any spectral parameter it can not
be employed in the nonlinear superposition (\ref{NLS}). However, taking $%
\phi _{0}=\phi _{\pm ,m,\alpha }^{\ell (1)}(x,t)$, $\phi _{1}=\phi _{\pm
,m,\beta }^{\ell (1)}(x,t)$ and $\phi _{2}=\phi _{\pm ,m,\gamma }^{\ell
(1)}(x,t)$ we identify from (\ref{BT}) the constants $\kappa _{1}=\pm m^{1/4}%
\limfunc{sn}\left( \alpha -\beta ,m\right) $ and $\kappa _{2}=\pm m^{1/4}%
\limfunc{sn}\left( \alpha -\gamma ,m\right) $, such that by (\ref{NLS}) we
obtain the new three soliton solutions 
\begin{equation}
\phi _{\pm ,\alpha \beta \gamma ,m}^{\ell (3)}=\phi _{\pm ,m,\alpha }^{\ell
(1)}+4\arctan \left[ \frac{\limfunc{sn}\left( \alpha -\beta ,m\right) +%
\limfunc{sn}\left( \alpha -\gamma ,m\right) }{\limfunc{sn}\left( \alpha
-\beta ,m\right) -\limfunc{sn}\left( \alpha -\gamma ,m\right) }\tan \left( 
\frac{\phi _{\pm ,m,\beta }^{\ell (1)}-\phi _{\pm ,m,\gamma }^{\ell (1)}}{4}%
\right) \right] .
\end{equation}%
As is most easily seen in the simpler solutions (\ref{phs1}) and (\ref{phs2}%
) the solutions for $\ell =>$ are also regular in the cases with spectral
parameter. We will not explore here how the limit $\beta ,\gamma \rightarrow
\alpha $ can be taken in these expressions and how one might iterate these
solutions in a similar way as explained in section 2. Instead we present the
same approach as in the previous sections.

\subsubsection{Degenerate cnoidal kink solutions}

Using the solution $\phi _{cn}^{(0)}$ as initial solutions and the solutions
(\ref{sp1}) and (\ref{sp4}) to the AKNS equations we are now in a position
to compute the degenerate cnoidal kink solutions using the Darboux
transformation involving Jordan states from%
\begin{equation}
\phi _{\pm ,m,\beta \beta }^{c\ell }=\phi _{cn}^{(0)}-2i\ln \frac{W\left[
\Phi _{\pm ,m,\beta }^{\ell },\partial _{\beta }\Phi _{\pm ,m,\beta }^{\ell }%
\right] }{W\left[ \Psi _{\pm ,m,\beta }^{\ell },\partial _{\beta }\Psi _{\pm
,m,\beta }^{\ell }\right] }.
\end{equation}%
A lengthy calculation yields%
\begin{equation}
\phi _{\pm ,m,\beta \beta }^{c\ell }=\phi _{cn}^{(0)}\pm 4\beta \kappa
-4\arctan \left[ i\frac{N_{\pm }^{\ell }-(N_{\pm }^{\ell })^{\ast }}{N_{\pm
}^{\ell }+(N_{\pm }^{\ell })^{\ast }}\right] .
\end{equation}%
where we defined the quantities%
\begin{eqnarray}
N_{\pm }^{<} &=&\Theta ^{2}\left( \frac{x-t}{2m^{1/4}}-\frac{i}{2}K^{\prime
}\right) \left\{ H^{2}\left( \frac{x-t}{2m^{1/4}}\pm \beta +\frac{i}{2}%
K^{\prime }\right) W_{\beta }\left[ \partial _{\beta }\Theta \left( \beta
\right) ,\Theta \left( \beta \right) \right] \right. \\
&&+\left. \Theta ^{2}\left( \beta \right) W_{\beta }\left[ H\left( \frac{x-t%
}{2m^{1/4}}\pm \beta +\frac{i}{2}K^{\prime }\right) ,\partial _{\beta
}H\left( \frac{x-t}{2m^{1/4}}\pm \beta +\frac{i}{2}K^{\prime }\right) \right]
\right\} ,~~~~\   \notag
\end{eqnarray}%
\begin{eqnarray}
N_{\pm }^{>} &=&\Theta ^{2}\left( \frac{x-t}{2m^{1/4}}-\frac{i}{2}K^{\prime
}\right) \left\{ \Theta ^{2}\left( \frac{x-t}{2m^{1/4}}\pm \beta +\frac{i}{2}%
K^{\prime }\right) W_{\beta }\left[ \partial _{\beta }H\left( \beta \right)
,H\left( \beta \right) \right] \right. \\
&&+\left. H^{2}\left( \beta \right) W_{\beta }\left[ \Theta \left( \frac{x-t%
}{2m^{1/4}}\pm \beta +\frac{i}{2}K^{\prime }\right) ,\partial _{\beta
}\Theta \left( \frac{x-t}{2m^{1/4}}\pm \beta +\frac{i}{2}K^{\prime }\right) %
\right] \right\} .~~~~~\   \notag
\end{eqnarray}%
Notice that the argument of the $\arctan $ is always real. These functions
are regular for real values of $\beta $. Furthermore we observe that the
additional speed spectral parameter is now separated from $x$ and $t$, so
that the degenerate solution has only one speed, i.e. the degenerate
solution is not displaced at any time. We depict this solution in figure \ref%
{degcniodal}.

\section{Degenerate multi-solitons from Hirota's direct method}

Finally we explore how the degenerate solutions may be obtained within the
context of Hirota's direct method \cite{hirota2004direct}. The key idea of
this solution procedure is to convert the original nonlinear equations into
bilinear forms, which can be solved systematically. When parameterizing $%
\phi (x,t)=2i\ln [g(x,t)/f(x,t)]$ the sine-Gordon equation (\ref{SG}) was
found \cite{HirotaSG,hirota2004direct} to be equivalent to the two equations%
\begin{equation}
D_{x}D_{t}f\cdot f+\frac{1}{2}(g^{2}-f^{2})=\lambda f^{2},\quad \text{%
and\quad }D_{x}D_{t}g\cdot g+\frac{1}{2}(f^{2}-g^{2})=\lambda g^{2},
\label{HirotaSG}
\end{equation}%
with $D_{x}$, $D_{t}$ denoting the Hirota derivatives. Explicitly we have $%
D_{x}D_{t}f\cdot f=2f^{2}(\ln f)_{xt}$. Taking $g=f^{\ast }$ the equations (%
\ref{HirotaSG}) become each others conjugate and with $\lambda =0$ can be
solved by the Wronskian 
\begin{equation}
f=W[\psi _{\alpha _{1}},\psi _{\alpha _{2}},\ldots ,\psi _{\alpha _{N}}],
\label{fW}
\end{equation}%
where%
\begin{equation}
\psi _{\alpha }=e^{\xi _{+}/2}+ic_{\alpha }e^{-\xi _{+}/2}.  \label{psi}
\end{equation}%
For simplicity we ignore here an overall constant that may be canceled out
without loss of generality and also do not treat the possibility $\xi
_{+}\rightarrow -\xi _{+}$ separately. This gives rise to the real valued $N$%
-soliton solutions%
\begin{equation}
\phi =2i\ln \frac{f^{\ast }}{f}=4\arctan \left( i\frac{f^{\ast }-f}{f^{\ast
}+f}\right) =4\arctan \frac{f_{i}}{f_{r}},  \label{dfg}
\end{equation}%
where $f=f_{r}+if_{i}$ with $f_{r},f_{i}\in \mathbb{R}$. For instance the
one, two and three-soliton solution obtained in this way are%
\begin{eqnarray}
\phi _{\alpha } &=&4\arctan (c_{\alpha }e^{-\xi _{+}^{\alpha }}), \\
\phi _{\alpha \beta } &=&4\arctan \left[ \Gamma _{\alpha \beta }\frac{%
c_{\beta }e^{\xi _{+}^{\beta }}-c_{\alpha }e^{\xi _{+}^{\alpha }}}{%
1+c_{\alpha }c_{\beta }e^{\xi _{+}^{\alpha }+\xi _{+}^{\beta }}}\right] , \\
\phi _{\alpha \beta \gamma } &=&4\arctan \left[ \frac{c_{\alpha }c_{\beta
}c_{\gamma }+c_{\alpha }\Gamma _{\alpha \beta }\Gamma _{\alpha \gamma
}e^{\xi _{+}^{\beta }+\xi _{+}^{\gamma }}+c_{\beta }\Gamma _{\beta \alpha
}\Gamma _{\beta \gamma }e^{\xi _{+}^{\alpha }+\xi _{+}^{\gamma }}+c_{\gamma
}\Gamma _{\gamma \alpha }\Gamma _{\gamma \beta }e^{\xi _{+}^{\alpha }+\xi
_{+}^{\beta }}}{c_{\beta }c_{\gamma }\Gamma _{\alpha \beta }\Gamma _{\alpha
\gamma }e^{\xi _{+}^{\alpha }}+c_{\alpha }c_{\gamma }\Gamma _{\beta \alpha
}\Gamma _{\beta \gamma }e^{\xi _{+}^{\beta }}+c_{\alpha }c_{\beta }\Gamma
_{\gamma \alpha }\Gamma _{\gamma \beta }e^{\xi _{+}^{\gamma }}+e^{\xi
_{+}^{\alpha }+\xi _{+}^{\beta }+\xi _{+}^{\gamma }}}\right] ,~~~~~~
\end{eqnarray}%
where $\Gamma _{xy}:=(x+y)/(x-y)$. We kept here the constants $c_{\alpha
},c_{\beta },c_{\gamma }$ generic as it was previously found \cite%
{CorreaFring} that they have be chosen in a specific way to render the
limits to the degenerate case finite.

Following the procedure outlined in \cite{CorreaFring} we replace the
standard solutions to the Schr\"{o}dinger equation in the non-degenerate
solution by Jordan states in the computation of $f$ in (\ref{fW}) as 
\begin{equation}
f=W[\psi _{\alpha },\partial _{\alpha }\psi _{\alpha },\partial _{\alpha
}^{2}\psi _{\alpha },\ldots ,\partial _{\alpha }^{N}\psi _{\alpha }].
\end{equation}%
We then recover from (\ref{dfg}) the degenerate kink solution $\phi _{\alpha
\alpha }$ and $\phi _{\alpha \alpha \alpha }$ in (\ref{deg2}) and (\ref{deg3}%
), respectively, with $c_{\alpha }=-1$ in (\ref{psi}). Unlike as in the
treatment of the Korteweg de-Vries equation \cite{CorreaFring} the equations
are already in a format that allows to carry out the limits $\lim_{\beta
\rightarrow \alpha }\phi _{\alpha \beta }=\phi _{\alpha \alpha }$ and $%
\lim_{\beta ,\gamma \rightarrow \alpha }\phi _{\alpha \beta \gamma }=\phi
_{\alpha \alpha \alpha }$ with the simple choices $c_{\alpha }=c_{\beta }=1$
and $c_{\alpha }=c_{\beta }=c_{\gamma }=-1$, respectively.

\section{Time-dependent displacements}

Let us now compute the time-dependent displacements by tracking the
one-soliton solution within a degenerate multi-soliton solution as explained
in \cite{cen2016time}.

\subsection{Time-dependent displacements for multi-kink solutions}

Unlike as for standard multi-soliton solutions one can not track the maxima
or minima for the kink-solutions as they might have maximal or minimal
amplitudes extending up to infinity. However, they have many intermediate
points in-between the extrema that are uniquely identifiable. For instance,
for the solutions constructed in sections 2. and 3.1 a suitable choice is
the point of inflection at half the maximal value, that is at $\phi
_{n\alpha }=\pi $ corresponding to $\tau _{n\alpha }=1$. For an $N$-soliton
solution $\phi _{N\alpha }~$with $N$ parameterized as $N=2n+1-\kappa $ we
find that these $N$ points are reached asymptotically%
\begin{equation}
\lim\limits_{t\rightarrow \infty }\tau _{2n+1-\kappa }\left( -\frac{t}{%
\alpha ^{2}}\pm \Delta _{n,\ell ,\kappa },t\right) =1,~~~~n=1,2,\ldots
;~\ell =\kappa ,\ldots ,n-1,n;~\kappa =0,1  \label{limit}
\end{equation}%
for the time-dependent displacements 
\begin{equation}
\Delta _{n,\ell ,\kappa }(t):=\frac{1}{\alpha }\ln \left[ \frac{(n-\ell )!}{%
(n+\ell -\kappa )!}\left( \frac{4t}{\alpha }\right) ^{2\ell -\kappa }\right] 
\label{delta}
\end{equation}%
For example, given a $5$-soliton solution $\phi _{5a}$ we have $n=2$, $%
\kappa =0~$and $\ell =0,1,2$, so that we can compare it with five laterally
displaced one-soliton solutions as depicted in figure \ref{displ}.

\FIGURE{ \epsfig{file=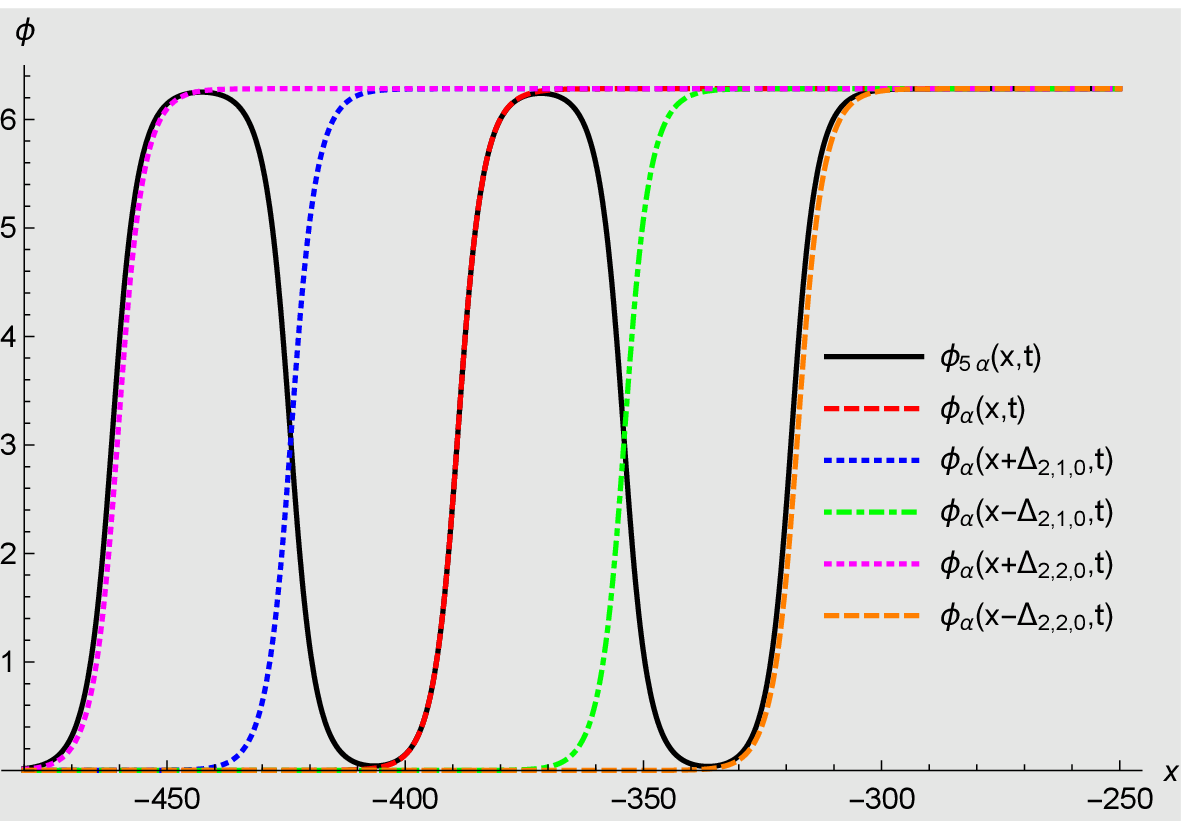,width=13.0cm} 
\caption{Degenerate 5-soliton solution compared with 5 time-dependently laterally displaced one-soliton solutions for $\alpha=0.3$ at time $t=35$.}
        \label{displ}}

Let us now derive this expression for the first examples. Introducing the
notation $x_{\pm }:=-t/\alpha ^{2}\pm 1/\alpha \ln \delta $, assuming that $%
\delta \sim t^{\mu }$, $\mu \geq 1$ and taking $y$ to be a polynomial in $t$%
, we obtain the useful auxiliary limits 
\begin{eqnarray}
\lim\limits_{t\rightarrow \infty }\lim\limits_{x\rightarrow x_{\pm }}\left(
y+\xi _{+}\right) &=&\lim\limits_{t\rightarrow \infty }\left( y\pm \ln
\delta \right) \approx \lim\limits_{t\rightarrow \infty }y, \\
\lim\limits_{t\rightarrow \infty }\lim\limits_{x\rightarrow x_{\pm }}\left(
y+\xi _{-}\right) &=&\lim\limits_{t\rightarrow \infty }\left( y+\frac{2t}{%
\alpha }\mp \ln \delta \right) \approx \lim\limits_{t\rightarrow \infty
}\left( y+\frac{2t}{\alpha }\right) , \\
\lim\limits_{t\rightarrow \infty }\lim\limits_{x\rightarrow x_{\pm }}\tau
_{1} &=&\lim\limits_{t\rightarrow \infty }\left( \delta ^{\pm 1}\right) .
\end{eqnarray}%
Using these expressions in (\ref{tau1}) - (\ref{tau5}) and the notation $%
\delta _{n,\ell ,\kappa }=\alpha \exp (\Delta _{n,\ell ,\kappa })$, $%
T=2t/\alpha $ we derive the asymptotic expressions for the $N$-soliton
solution for the lowest values of $N$ 
\begin{eqnarray*}
\lim\limits_{\substack{ t\rightarrow \infty  \\ x\rightarrow x_{\pm }}}\tau
_{2} &\approx &\lim\limits_{t\rightarrow \infty }\frac{2T\delta ^{\pm }}{%
1+(\delta ^{\pm })^{2}}=\lim\limits_{t\rightarrow \infty }\frac{2T}{\delta
^{\pm }}=1~~\text{for }\delta ^{\pm }=2T=\delta _{1,1,1}, \\
\lim\limits_{\substack{ t\rightarrow \infty  \\ x\rightarrow x_{\pm }}}\tau
_{3} &\approx &\lim\limits_{t\rightarrow \infty }\frac{2T^{2}+(\delta ^{\pm
})^{2}}{2T^{2}\delta ^{\pm }}=1~~~~~~~\ ~\text{for }\delta ^{\pm
}=2T^{2}=\delta _{1,1,0}\text{ or~}\delta ^{\pm }=1, \\
\lim\limits_{\substack{ t\rightarrow \infty  \\ x\rightarrow x_{\pm }}}\tau
_{4} &\approx &\lim\limits_{t\rightarrow \infty }\frac{4T^{3}\left[
1+(\delta ^{\pm })^{2}\right] }{4T^{4}\delta ^{\pm }+3(\delta ^{\pm })^{3}}%
=1~~\text{for }\delta ^{\pm }=T=\delta _{2,1,1}\text{ or~}\delta ^{\pm }=%
\frac{4}{3}T^{3}=\delta _{2,2,1}, \\
\lim\limits_{\substack{ t\rightarrow \infty  \\ x\rightarrow x_{\pm }}}\tau
_{5} &\approx &\lim\limits_{t\rightarrow \infty }\frac{4T^{6}\delta ^{\pm
}+9(\delta ^{\pm })^{3}}{4T^{6}+6T^{4}(\delta ^{\pm })^{2}}=1~~\text{for }%
\delta ^{\pm }=\frac{2}{3}T^{4}=\delta _{2,2,0}~\text{or~}\delta ^{\pm }=%
\frac{2}{3}T^{2}=\delta _{2,1,0}\text{ or~}\delta ^{\pm }=1,~~~
\end{eqnarray*}%
The limits need to be carried out in consecutive order, i.e. first replace $%
x\rightarrow x_{\pm }$ and then compute $t\rightarrow \infty $. These are
the first explicit examples for the asymptotic values all confirming (\ref%
{delta}). Similarly we have computed examples for higher values of $N$ that
may also be cast into the general formula (\ref{delta}). So far we have not
obtained a generic proof valid for any $N$.

Of course one may easily convert the shifts from light-cone to the original
variables. Having computed the lateral displacement $\Delta _{x}$ the
time-displacement is obtained as usual from $\Delta _{t}=-\Delta _{x}/v$,
where $v=1/\alpha ^{2}$ in our case. Then one simply obtains $\Delta
_{X}=\Delta _{x}(1-\alpha ^{2})$ and $\Delta _{T}=\Delta _{x}(1+\alpha ^{2})$%
.

\subsection{Time-dependent displacements for breather solutions}

For the breathers it is even less evident what point in the solution is
suitable for tracking due to the overall oscillation. However, since we are
only interested in the net movement we can neglect the internal oscillation
and determine the displacement for an enveloping function that surrounds the
breather and moves with the same overall speed. For the one-breather
solution an enveloping function is obtained by setting the $\sin $-function
in (\ref{b1}) to $1$, obtaining%
\begin{equation}
\phi _{\tilde{\alpha},1/\tilde{\alpha}}^{\text{env}}=4\arctan \left[ \frac{%
\theta }{\cosh \left[ \theta (t+x)/\sqrt{1+\theta ^{2}}\right] }\right] .
\end{equation}%
This function is depicted together with the breather solution in figure \ref%
{breatherenv} having a clearly identifiable maximum at $4\arctan \theta $
which we can track.

\FIGURE{ \epsfig{file=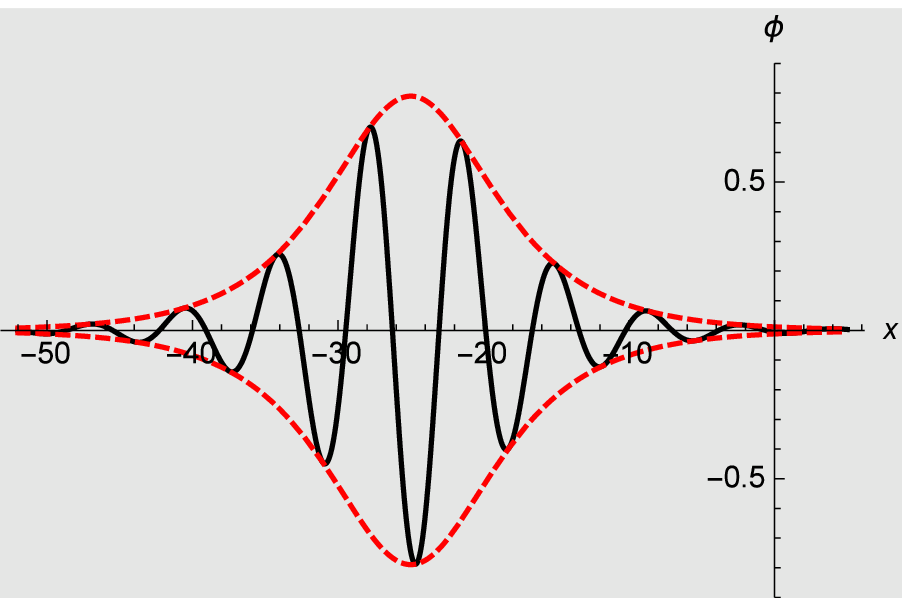,width=7.3cm} \epsfig{file=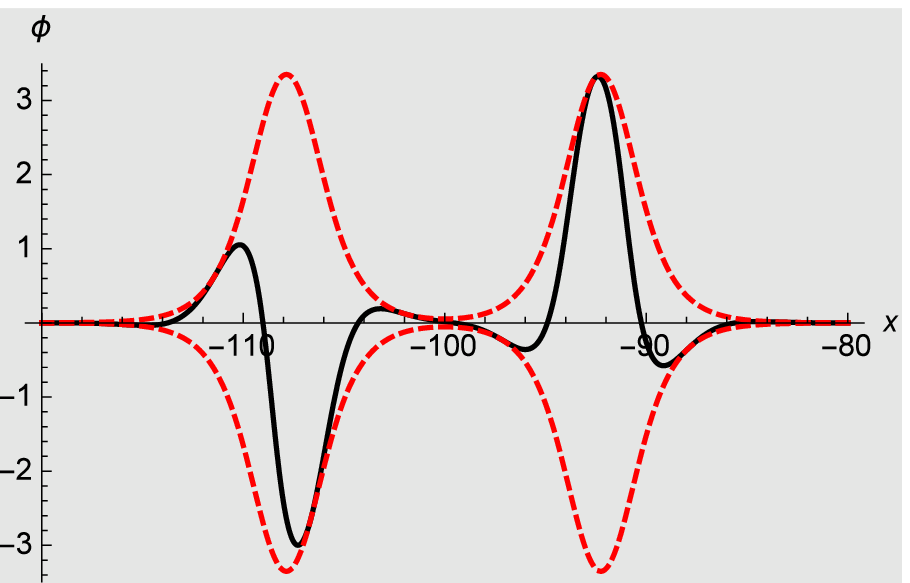,width=7.3cm}
\caption{One-breather solution surrounded by enveloping function $\pm \phi _{\tilde{\alpha},1/\tilde{\alpha}}^{\text{env}}$ at $t=25$ for $\theta=1/5$ panel (a) 
 and degenerate two-breather solution surrounded by enveloping function $\pm \phi _{\tilde{\alpha},\tilde{\alpha},1/\tilde{\alpha},1/\tilde{\alpha}}^{\text{env}}$ at $t=100$ for $\theta=4/3$ panel (b).}
        \label{breatherenv}}

We compare this now with the breather solution $\phi _{\tilde{\alpha},\tilde{%
\alpha},1/\tilde{\alpha},1/\tilde{\alpha}}$ constructed in section 3.1.3.
Taking for that solution $\sin \left[ (t-x)/\sqrt{1+\theta ^{2}}\right]
\rightarrow 0$ and $\cos \left[ (t-x)/\sqrt{1+\theta ^{2}}\right]
\rightarrow 1$ we obtain the enveloping function%
\begin{equation}
\phi _{\tilde{\alpha},\tilde{\alpha},1/\tilde{\alpha},1/\tilde{\alpha}}^{%
\text{env}}=4\arctan \left[ \frac{4(t-x)\theta \sqrt{1+\theta ^{2}}\cosh
\left( \frac{(t+x)\theta }{\sqrt{1+\theta ^{2}}}\right) }{2\theta
^{2}(t-x)^{2}+2(t+x)^{2}+1+\theta ^{-2}+(1+\theta ^{-2})\cosh \left( \frac{%
2\theta (t+x)}{\sqrt{1+\theta ^{2}}}\right) }\right] .
\end{equation}%
This function tends asymptotically to the maximal value of the one-breather
enveloping function 
\begin{equation}
\lim\limits_{t\rightarrow \infty }\phi _{\tilde{\alpha},\tilde{\alpha},1/%
\tilde{\alpha},1/\tilde{\alpha}}^{\text{env}}\left( -t\pm \Delta _{\tilde{%
\alpha},\tilde{\alpha},1/\tilde{\alpha},1/\tilde{\alpha}},t\right) =4\arctan
\left( \theta \right) ,
\end{equation}%
when shifted appropriately with the time-dependent displacement%
\begin{equation}
\Delta _{\tilde{\alpha},\tilde{\alpha},1/\tilde{\alpha},1/\tilde{\alpha}}(t)=%
\frac{1}{\theta }\sqrt{1+\theta ^{2}}\ln \left( \frac{4\theta ^{2}t}{\sqrt{%
1+\theta ^{2}}}\right) .
\end{equation}%
Similarly we may compute the displacements for the solutions $\phi _{\tilde{%
\alpha},\tilde{\alpha},\tilde{\alpha},1/\tilde{\alpha},1/\tilde{\alpha},1/%
\tilde{\alpha}}$ etc.

\section{Conclusions}

We have constructed various types of degenerate multi-soliton solutions from
three different methods commonly used in the context of classical nonlinear
integrable systems. Using the recurrence relations constructed in section 2
from B\"{a}cklund transformations is the most efficient way to obtain $N$%
-soliton solutions for large values of $N$. These equations are easily
implemented in computer calculations. By just requiring a simple solution to
the original nonlinear equation they also have a relatively easy starting
point. However, the equations are less universal as those presented in
section 3 using Jordan states in Darboux-Crum transformations. The
disadvantage of this method is that they need in addition the solutions to
the AKNS equations. For large values of $N$ the computations become more
involved than the recurrence relations for the B\"{a}cklund transformations.
Finally, Hirota's direct method turned out to be the simplest approach for
the sine-Gordon model as the limit to the degenerate case could be taken
directly from the $N$-soliton solutions with different spectral parameters.
However, as discussed for the Korteweg de-Vries equation in \cite%
{CorreaFring} this is not a general feature and one might have to tune the
arbitrary constants involved in a very specific and nontrivial way,
especially when one wishes to implement shift parameters.

We computed the explicit analytic expression for the asymptotic
time-dependent displacements between the one-soliton constituents. It turns
out that formula (\ref{uni}) is shared by multi-kink solutions in the
sine-Gordon equation and the multi-soliton solutions of the Korteweg
de-Vries equation\ when expression part of the combinations of the spectral
parameter in terms of the appropriate speed in the model. It seems obvious
to conjecture that this might also hold for other models, which would be
interesting to investigate. For the breather solutions we computed the
displacements for the enveloping functions. Interestingly we also found that
it is possible to construct compound solutions that travel uniformly and do
not display any displacement at any time.

\medskip

\noindent \textbf{Acknowledgments:} JC is supported by a City, University of
London Research Fellowship. FC was partially supported by Fondecyt grant
1171475 and the Alexander von Humboldt Foundation.

\newif\ifabfull\abfulltrue


\newif\ifabfull\abfulltrue

\begin{appendix}
\section{Appendix}
In this appendix we present some derivations of identities used in the manuscript.

First we present a derivation of identity (\ref{Id1}). We start by
considering the limit for $n=2$%
\begin{equation}
\lim_{\beta \rightarrow \alpha }\frac{\alpha +\beta }{\alpha -\beta }\tan
\left( \frac{\phi _{\beta }-\phi _{\alpha }}{4}\right)   \label{a1}
\end{equation}%
by taking $\beta =\alpha +h~$and letting $h$ tend to zero. So (\ref{a1}) may
be written as%
\begin{eqnarray}
\lim_{h\rightarrow 0}\frac{2\alpha +h}{-h}\tan \left( \frac{\phi _{\alpha
+h}-\phi _{\alpha }}{4}\right)  &=&-2\alpha \lim_{h\rightarrow 0}\frac{1}{%
\cos \left[ \frac{\phi _{\alpha +h}-\phi _{\alpha }}{4}\right] }\frac{\sin %
\left[ \frac{\phi _{\alpha +h}-\phi _{\alpha }}{4}\right] }{h},  \notag \\
&=&-2\alpha \lim_{h\rightarrow 0}\frac{\phi _{\alpha +h}-\phi _{\alpha }}{4h}%
, \\
&=&-\frac{\alpha }{2}\frac{d\phi _{\alpha }}{d\alpha }.  \notag
\end{eqnarray}%
Viewing $\phi $ as a function of $\alpha $ we identified in the last
equality the standard expression for the derivative. Using this expression
the non-degenerate and degenerate two-soliton solutions mav be written as%
\begin{equation}
\phi _{\alpha \beta }=4\arctan \left[ \frac{\alpha +\beta }{\alpha -\beta }%
\tan \left( \frac{\phi _{\beta }-\phi _{\alpha }}{4}\right) \right] ~,~~~\ 
\text{and~\ ~~\ ~}\phi _{\alpha \alpha }=-4\arctan \left[ \frac{\alpha }{2}%
\frac{d\phi _{\alpha }}{d\alpha }\right] ,  \label{ndegdeg}
\end{equation}%
respectively.

Next we considering the limit for $n=3$%
\begin{equation}
\lim_{\beta \rightarrow \alpha }\frac{\alpha +\beta }{\alpha -\beta }\tan
\left( \frac{\phi _{\alpha \beta }-\phi _{\alpha \alpha }}{4}\right) 
\end{equation}%
We re-write this expression as%
\begin{equation}
\lim_{\beta \rightarrow \alpha }\lim_{\gamma \rightarrow \alpha }\frac{\beta
+\gamma }{\beta -\gamma }\tan \left( \frac{\phi _{\alpha \gamma }-\phi
_{\alpha \beta }}{4}\right) =\lim_{\beta \rightarrow \alpha }\lim_{\gamma
\rightarrow \beta }\frac{\beta +\gamma }{\beta -\gamma }\tan \left( \frac{%
\phi _{\alpha \gamma }-\phi _{\alpha \beta }}{4}\right) ,
\end{equation}%
which when setting $\gamma =\beta +h$ becomes%
\begin{eqnarray}
\lim_{\beta \rightarrow \alpha }\lim_{h\rightarrow 0}\frac{2\beta +h}{-h}%
\tan \left( \frac{\phi _{\alpha \beta +h}-\phi _{\alpha \beta }}{4}\right) 
&=&-2\lim_{\beta \rightarrow \alpha }\beta \lim_{h\rightarrow 0}\frac{1}{%
\cos \left[ \frac{\phi _{\alpha \beta +h}-\phi _{\alpha \beta }}{4}\right] }%
\frac{\sin \left[ \frac{\phi _{\alpha \beta +h}-\phi _{\alpha \beta }}{4}%
\right] }{h}  \notag \\
&=&-2\lim_{\beta \rightarrow \alpha }\beta \lim_{h\rightarrow 0}\frac{\phi
_{\alpha \beta +h}-\phi _{\alpha \beta }}{4h} \\
&=&-\frac{\alpha }{2}\lim_{\beta \rightarrow \alpha }\frac{d\phi _{\alpha
\beta }}{d\beta }  \notag \\
&=&-\frac{\alpha }{4}\frac{d\phi _{\alpha \alpha }}{d\alpha }  \notag
\end{eqnarray}%
The last equality follows from a direct calculation using the ecpressions in
(\ref{ndegdeg}). We compute%
\begin{equation}
\frac{1}{2}\frac{d\phi _{\alpha \alpha }}{d\alpha }=-\frac{\frac{d\phi
_{\alpha }}{d\alpha }+\alpha \frac{d^{2}\phi _{\alpha }}{d\alpha ^{2}}}{%
1+(\alpha /2)^{2}\left( \frac{d\phi _{\alpha }}{d\alpha }\right) ^{2}}
\label{tt}
\end{equation}%
and%
\begin{equation}
\lim_{\beta \rightarrow \alpha }\frac{d\phi _{\alpha \beta }}{d\beta }=\frac{%
32\alpha (\phi _{\beta }-\phi _{\alpha })+16(\alpha ^{2}-\beta ^{2})\frac{%
d\phi _{\beta }}{d\beta }}{16(\alpha -\beta )^{2}+(\alpha +\beta )^{2}(\phi
_{\alpha }-\phi _{\beta })^{2}}.
\end{equation}%
Replacing $\phi _{\beta }\rightarrow \phi _{\alpha }+h\frac{d\phi _{\alpha }%
}{d\alpha }+h^{2}/2\frac{d^{2}\phi _{\alpha }}{d\alpha ^{2}}$,$~$ $\beta
\rightarrow \alpha +h$ and identifying the derivitives as above, this
becomes precisely the right had side of (\ref{tt}). It is now clear how to
proceed for larger $n$. We have verified the identity up to $n=6$. 

Identity (\ref{axt}) is a simple variable transformation based on the
assumption that $\phi _{n\alpha }(x,t)$ can alway be expressed as $\phi
_{n\alpha }(\xi _{+},\xi _{-})$. So we compute%
\begin{eqnarray}
\frac{\partial \phi _{n\alpha }}{\partial x} &=&\alpha \frac{\partial \phi
_{n\alpha }}{\partial \xi _{+}}-\alpha \frac{\partial \phi _{n\alpha }}{%
\partial \xi _{-}},  \label{i1} \\
\frac{\partial \phi _{n\alpha }}{\partial t} &=&\frac{1}{\alpha }\frac{%
\partial \phi _{n\alpha }}{\partial \xi _{+}}+\frac{1}{\alpha }\frac{%
\partial \phi _{n\alpha }}{\partial \xi _{-}},  \label{i2} \\
\frac{\partial \phi _{n\alpha }}{\partial \alpha } &=&(x-\frac{t}{\alpha ^{2}%
})\frac{\partial \phi _{n\alpha }}{\partial \xi _{+}}-(x+\frac{t}{\alpha ^{2}%
})\frac{\partial \phi _{n\alpha }}{\partial \xi _{-}}.  \label{i3}
\end{eqnarray}%
Comparing (\ref{i1}), (\ref{i2}) and (\ref{i3}) we can eliminate the
derivtives with respect to $\xi _{+}$, $\xi _{-}$ and obtain (\ref{axt}).

\end{appendix}

\end{document}